\RequirePackage{fix-cm}
\documentclass[
pdftex,
twocolumn,
final,
epjc3]{svjour3}  

\smartqed
\journalname{Eur. Phys. J. C}

\RequirePackage{xspace}
\RequirePackage{graphicx}
\RequirePackage{amsmath}
\RequirePackage{mathtools}
\RequirePackage{booktabs}
\RequirePackage[tableposition=bottom]{caption}
\RequirePackage{hyperref}
\RequirePackage{nicefrac}
\RequirePackage{cite}
\RequirePackage[a4paper, total={6.5in, 8.5in}]{geometry}
\RequirePackage{xcolor}
\RequirePackage{amssymb}
\RequirePackage{afterpage}
\RequirePackage[utf8]{inputenc}
\RequirePackage{enumitem}
\RequirePackage{calc}
\RequirePackage{booktabs}
\RequirePackage{multirow}
\RequirePackage{mathtools}
\RequirePackage{cuted}
\RequirePackage{etoolbox}
\RequirePackage{esint}
\RequirePackage{verbatim}
\RequirePackage{hepunits}
\RequirePackage{wasysym}

\usepackage{definitions}

\def\figWidth{\linewidth}

\begin{document}

\title{Phenomenological assessment of proton mechanical properties from deeply virtual Compton scattering}

\author{
H.~Dutrieux\thanksref{email0,address2}
\and
C.~Lorc\'e\thanksref{email1,address1}
\and
H.~Moutarde\thanksref{email2,address2}
\and
P.~Sznajder\thanksref{email3,address3}
\and
A.~Trawi\'nski\thanksref{email4, address2, address1} 
\and
J.~Wagner\thanksref{email5,address3} 
}

\thankstext{email0}{e-mail: herve.dutrieux@cea.fr}
\thankstext{email1}{e-mail: cedric.lorce@polytechnique.edu}
\thankstext{email2}{e-mail: herve.moutarde@cea.fr}
\thankstext{email3}{e-mail: pawel.sznajder@ncbj.gov.pl}
\thankstext{email4}{e-mail: arkadiusz.trawinski@gmail.com}
\thankstext{email5}{e-mail: jakub.wagner@ncbj.gov.pl}


\institute{%
IRFU, CEA, Universit\'e Paris-Saclay, F-91191 Gif-sur-Yvette, France \label{address2}
\and
CPHT, CNRS, Ecole Polytechnique, Institut Polytechnique de Paris, Route de Saclay, 91128 Palaiseau, France \label{address1}
\and
National Centre for Nuclear Research (NCBJ), Pasteura 7, 02-093 Warsaw, Poland \label{address3}
}

\date{Received: date / Accepted: date}

\maketitle

\sloppy

\begin{abstract}
A unique feature of generalised parton distributions is their relation to the QCD energy-momentum tensor. In particular, they provide access to the mechanical properties of the proton \ie the distributions of pressure and shear stress induced by its quark and gluon structure. In principle the pressure distribution can be experimentally determined in a model-independent way from a dispersive analysis of deeply virtual Compton scattering data through the measurement of the subtraction constant. In practice the kinematic coverage and accuracy of existing experimental data make this endeavour a challenge. Elaborating on recent global fits of deeply virtual Compton scattering measurements using artificial neural networks, our analysis presents the current knowledge on this subtraction constant and assesses the impact of the most frequent systematic assumptions made in this field of research. This study will pave the way for future works when more precise data will become available, \eg obtained in the foreseen electron-ion colliders EIC and EIcC. 
\end{abstract}

\keywords{3D Nucleon Structure \and Nucleon Tomography \and Global Fit \and Deeply Virtual Compton Scattering \and DVCS \and Compton Form Factor \and CFF \and Dispersion Relation \and Subtraction Constant \and Generalised Parton Distribution \and GPD \and Artificial Neural Network \and Genetic Algorithm \and EIC \and EicC \and Jefferson Lab \and PARTONS Framework}
\PACS{12.38.-t \and 13.60.-r \and 13.60.Fz \and 14.20.-c}

%
%
\section{Introduction}
\label{sec:introduction}

Quantum chromodynamics (QCD) provides today's standard description of hadrons. In this theory, hadrons are depicted as complex objects made out of fundamental building blocks referred to as partons: quarks and gluons. This picture has been extensively studied over the last fifty years, but many of the basic questions about partonic media still await a precise quantitative answer to this day. In particular, in spite of important theoretical advances since the beginning of the 2000s, our knowledge of distributions of certain properties like pressure and shear stress inside hadrons -- the so-called ``mechanical'' properties -- is still poor. Understanding those properties therefore remains among the main challenges faced by modern nuclear and high energy physics.

QCD factorisation theorems provide us with tools to study the properties of partonic media up to an arbitrarily high accuracy, only limited by the order of the perturbative expansion describing the short distance part of scattering processes. In particular, generalised parton distributions (GPDs) offer a rigorous theoretical framework that can be used to study the 3D structure of hadrons \cite{Mueller:1998fv, Ji:1996ek, Ji:1996nm, Radyushkin:1996ru, Radyushkin:1997ki}. GPDs offer a consistent unified framework encompassing the usual parton distribution functions (PDFs) and elastic form factors (EFFs), and providing even more information. Such a non-trivial connection is used to obtain tomographic pictures of the nucleon, where spatial distributions of partons carrying a fraction of the nucleon momentum are projected onto the plane perpendicular to the direction of the nucleon motion \cite{Burkardt:2000za, Burkardt:2002hr, Burkardt:2004bv}. Another exciting feature of GPDs is their relation to the QCD energy-momentum tensor (EMT), which would otherwise probably only be accessible via graviton scattering. This unique relation allows to evaluate the total angular momentum carried by gluons or quarks of a given flavour \cite{Ji:1996ek, Ji:1996nm}, which is essential to solve the long-standing puzzle of the nucleon spin decomposition that emerged 30 years ago with the EMC measurements \cite{Ashman:1987hv}. On top of this, the relation between GPDs and the EMT can also be used to access information about the mechanical properties of partonic systems \cite{Goeke:2007fp, Polyakov:2018zvc}, like distributions of pressure inside the nucleon -- which is the subject of this article. 

The potential of studying the mechanical properties of the proton in the formalism of GPDs was first recognised in Ref.~\cite{Polyakov:2002yz}. This subject has recently gained a lot of interest and seen vital theoretical progress. In addition, precise data have become available, recently allowing for the first data-driven estimates of a nucleon pressure distribution \cite{Burkert:2018bqq, Kumericki:2019ddg}. These analyses aim at extracting the so-called subtraction constant from deeply virtual Compton scattering (DVCS) data. The constant naturally appears in the dispersive studies of DVCS amplitudes and it carries information about one of the EMT form factors that is crucial for the understanding of the mechanical properties. We point out that the subject in general has been recognised as a key element of current or future experiments, in particular those to be conducted in the electron-ion collider (EIC) \cite{Accardi:2012qut}, Chinese electron-ion collider (EIcC) \cite{Chen:2018wyz}, large hadron-electron collider (LHeC) \cite{AbelleiraFernandez:2012cc} and high-intensity heavy ion accelerator facility (HIAF) \cite{Yang:2013yeb}.

In this article we benchmark the extraction of the subtraction constant from DVCS data. We present results of our extraction of the subtraction constant from the existing world DVCS proton data. To deliver a self-contained exposition, we remind the leading-order (LO) QCD evolution formulas, which in general are known, but appear in the literature in a scattered form. Compared to the first data-driven estimates, our analysis is characterised by a reduced model dependence, which is achieved by the use of artificial neural network techniques to describe the contribution of the four chiral-even leading-twist GPDs. The results of this analysis summarise our current knowledge of the DVCS subtraction constant and assess the impact of the most common systematic assumptions made in that research topic. This will pave the way for future studies when new precise data will be released. 

This article is organised as follows. A brief introduction to the theoretical frameworks of EMT and GPDs is given in Sects.~\ref{sec:emt} and \ref{sec:formalism} respectively. Section~\ref{sec:evolution} is dedicated to the LO evolution of the subtraction constant, and Sect.~\ref{sec:model} to its modelling. The extraction based on the artificial neural network technique is presented in Sect.~\ref{sec:fits} and the corresponding results are discussed in Sect.~\ref{sec:results}. The summary is given in Sect.~\ref{sec:summary}.

\paragraph{Notations} Throughout the text we denote  spatial vectors by boldface symbols. $\eta_{\mu\nu} = \text{diag}(+,-,-,-)$ is the Minkowski metric. For convenience, we use the compact notations: $a^{\{\mu}b^{\nu\}}  = a^\mu b^\nu + a^\nu b^\mu$ and $a^{[\mu}b^{\nu]} = a^\mu b^\nu - a^\nu b^\mu$.

%
%
\section{Energy-momentum tensor}
\label{sec:emt}

In the most general case, the proton matrix elements of the local gauge-invariant EMT operator can be parameterised in terms of five gravitational form factors (GFFs): $A_{a}(t)$, $B_{a}(t)$, $C_{a}(t)$, $\bar C_{a}(t)$ and $D_{a}(t)$; as follows~\cite{Bakker:2004ib,Leader:2013jra}:
\begin{strip}
\makeatletter
\setbool{@fleqn}{false}
\makeatother
\begin{align}
\label{eq:TmunuUU}
\bra{p',\uvec s'} T^{\mu\nu}_a(0)\ket{p,\uvec s}
= \bar u(p',\uvec s') \Bigg\{&
\frac{P^\mu P^\nu}{M}\,A_a(t)
+ \frac{\Delta^\mu\Delta^\nu - \eta^{\mu\nu}\Delta^2}{M}\, C_a(t)
+ M \eta^{\mu\nu}\bar C_a(t) \nn
\\
&+ \frac{P^{\{\mu} i\sigma^{\nu\}\rho}\Delta_\rho}{4M}\left[A_a(t)+B_a(t)\right]+ \frac{P^{[\mu} i\sigma^{\nu]\rho}\Delta_\rho}{4M}\,D_a(t)\Bigg\} u(p,\uvec s) \,.
\end{align}
\makeatletter
\setbool{@fleqn}{true}
\makeatother
\end{strip}
Here, $t=\Delta^2$ with $\Delta=p'-p$ the four-momentum transfer to the proton, $P=(p'+p)/2$ is the average four-momentum, $M$ is the proton mass, $u(p,\uvec s)$, $\bar u(p',\uvec s')$ are Dirac spinors with the covariant normalisation $\bar u(p,\uvec s) u(p,\uvec s) = 2M$, and $\uvec{s}$, $\uvec{s'}$ are the rest-frame polarisation vectors. The label $a$ denotes either the quark flavour ($a=q$) or the gluon ($a=g$) contribution to the EMT. We refer to Ref.~\cite{Lorce:2015lna} for a detailed discussion of the specific expressions of the EMT, and their decompositions into quark and gluon contributions. 

In the following, a summation over parton types $a$ is meant each time a generic GFF $G \in \{A, B, C, \bar C, D\}$ is written without any parton type index:
\begin{equation}
G(t) = \sum_{a=q,g} G_a(t)  \;, \label{def-gff-G-noindex} 
\end{equation}
and similarly a summation over quark flavours $q$ (flavour-singlet combination) is meant each time this generic GFF $G$ is written with the superscript $S$:
\begin{equation}
G^S(t) = \sum_{a=q} G_a(t)  \;. \label{def-gff-G-singlet}
\end{equation}
When needed, we indicate the quark content with:
\begin{eqnarray}
G^{u+d}(t) & = & G^u(t) + G^d(t) \;, \\
G^{u+d+s}(t) & = & G^u(t) + G^d(t) + G^s(t) \;.
\end{eqnarray}

Poincar\'e symmetry implies the following constraints on the GFFs~\cite{Ji:1996ek,Brodsky:2000ii,Lowdon:2017idv,Lorce:2019sbq}:
\begin{align}
A(0) &= 1 \,, \\
B(0) &= 0 \,, \\
\bar C(t) &= 0 \,, 
\end{align}
and~\cite{Lorce:2017wkb}:
\begin{align}
D^S(t) &= -G_A(t) \,, \\
D_g(t) &= 0 \,,
\end{align}
where $G_A(t)$ is the flavour-singlet axial-vector form factor. The GFFs $A_a(t)$, $B_a(t)$ and $C_a(t)$ can be related to leading-twist GPDs and thus can be probed at present experimental facilities, see Sect.~\ref{sec:formalism}. Accessing $\bar C_a(t)$ is more involved since it is related to higher-twist distributions \cite{Leader:2012ar,Leader:2013jra,Tanaka:2018wea}.

The static EMT is defined as the following Fourier transform in the Breit frame, where $\boldsymbol P=\boldsymbol 0$ and $t=-\boldsymbol\Delta^2$~\cite{Polyakov:2002yz,Polyakov:2018zvc,Lorce:2018egm}:
\begin{equation}
\label{eq:TFT}
\mathcal T^{\mu\nu}_a(\boldsymbol r)=\int\frac{\ud^3\boldsymbol\Delta}{(2\pi)^3}\,e^{-i\boldsymbol\Delta\cdot\boldsymbol r}\,\frac{\bra{p',\uvec s} T^{\mu\nu}_a(0)\ket{p,\uvec s}}{2P^0} \,.
\end{equation} 
It indicates how energy and momentum are distributed inside the proton with the canonical polarisation $\uvec s$. Writing $r = |\boldsymbol r|$ the radial coordinate, the energy $\varepsilon_a(r)$, radial pressure $p_{r,a}(r)$ and tangential pressure $p_{t,a}(r)$ distributions can be expressed in the Breit frame as Fourier transforms of the GFFs $A_a(t)$, $B_a(t)$, $C_a(t)$ and $\bar C_a(t)$:
\begin{align}
\label{eq:epsilonbis}
\varepsilon_a(r)
&=M\int\frac{\ud^3\boldsymbol\Delta}{(2\pi)^3}\,e^{-i\boldsymbol\Delta\cdot\boldsymbol r} 
\nn\\
&\times \left\{A_a(t)+\bar{C}_a(t) + \frac{t}{4M^2}\Big[B_a(t)- 4C_a(t)\Big]\right\}\,,
\\
\label{eq:prbis}
p_{r,a}(r)
&= M\int\frac{\ud^3\boldsymbol\Delta}{(2\pi)^3}\,e^{-i\boldsymbol\Delta\cdot\boldsymbol r}
\nn\\
&\times \left\{-\bar C_a(t)-\frac{4}{r^2}\frac{t^{-1/2}}{M^2}\frac{\ud}{\ud t}\!\Big[t^{3/2}\,C_a(t)\Big]\right\} \,,
\\
\label{eq:ptbis}
p_{t,a}(r)
&= M\int\frac{\ud^3\boldsymbol\Delta}{(2\pi)^3}\,e^{-i\boldsymbol\Delta\cdot\boldsymbol r}
\nn\\
&\times \left\{-\bar C_a(t)+\frac{4}{r^2}\frac{t^{-1/2}}{M^2}\frac{\ud}{\ud t}\!\left(t\frac{\ud}{\ud t}\!\left[t^{3/2}\,C_a(t)\right]\right)\right\} \,.
\end{align}

In a relativistic system like the proton, pressure forces are usually not isotropic. The isotropic pressure $p_a(r)$ and pressure anisotropy $s_a(r)$ are then defined in terms of the radial and tangential pressures: 
\begin{align}
\label{eq:def-isotropic-pressure}
p_a(r) & =  \big[p_{r,a}(r)+2p_{t,a}(r)\big]/3 \;, \\
\label{eq:def-pressure-anisotropy}
s_a(r) & = p_{r,a}(r)-p_{t,a}(r) \;,
\end{align}
or, in terms of GFFs:
\begin{align}
\label{eq:pbis}
p_a(r)
&= M\int\frac{\ud^3\boldsymbol\Delta}{(2\pi)^3}\,e^{-i\boldsymbol\Delta\cdot\boldsymbol r}
\left\{-\bar{C}_a(t) +\frac{2}{3} \frac{t}{M^2}\,C_a(t)\right\} \,,
\\
\label{eq:sbis}
s_a(r)
&= -\frac{4 M}{r^2}\int\frac{\ud^3\boldsymbol\Delta}{(2\pi)^3}\,
e^{-i\boldsymbol\Delta\cdot\boldsymbol r}\,\frac{t^{-1/2}}{M^2}\frac{\ud^2}{\ud t^2}\!\left[t^{5/2}\,C_a(t)\right] \,.
\end{align}
Among the five distributions: $\varepsilon_a(r)$, $p_{r,a}(r)$, $p_{t,a}(r)$, $p_a(r)$ and $s_a(r)$; the pressure anisotropy $s_a(r)$ is the only one that does not depend on the GFF $\bar C_a(t)$. This feature makes its current experimental access the least challenging.

These distributions permit the introduction of additional definitions of proton radius beyond the usual electromagnetic charge radii. Indeed the distributions $\varepsilon(r)=\sum_a\varepsilon_a(r)$ and $p_r(r)=\sum_ap_{r,a}(r)$ are expected to be positive and can be used to define the energy (or mass) and mechanical radii of the nucleon~\cite{Polyakov:2018zvc, Lorce:2018egm} in terms of the GFFs $A$ and $C$ (summed over all constituents):
\begin{align}
\langle r^2\rangle_E&=\frac{1}{M}\int\ud^3\boldsymbol r\,r^2\,\varepsilon(r)=6\left[A'(0)-\frac{C(0)}{M^2}\right], \label{eq:def-energy-radius} \\
\langle r^2\rangle_\text{mech}&=\frac{1}{\mathcal P_r}\int\ud^3\boldsymbol r\,r^2\,p_r(r)=\frac{6C(0)}{\int_{-\infty}^0\ud t\,C(t)} \;, \label{eq:def-mechanical-radius}
\end{align}
with $\mathcal P_r=\int\ud^3\boldsymbol r\,p_r(r)$ and $A'(t)=\ud A(t)/\ud t$.

%
%
\section{Generalised parton distributions}
\label{sec:formalism}

GPDs have been introduced by the end of the 90s \cite{Mueller:1998fv, Ji:1996ek, Ji:1996nm, Radyushkin:1996ru, Radyushkin:1997ki} and now constitute a mature field of research~\cite{Goeke:2001tz, Diehl:2003ny, Belitsky:2005qn, Boffi:2007yc, Guidal:2013rya, Mueller:2014hsa} with experimental programmes carried out at CERN (COMPASS), DESY (HERMES, H1 and ZEUS), Jefferson Lab (Halls A, B and C) and in the future at the electron-ion collider EIC. References \cite{Favart:2015umi, Kumericki:2016ehc, dHose:2016mda, Grocholski:2019pqj} provide a recent account of the experimental and phenomenological status and prospects of three exclusive channels which attract most of current experimental interest, namely DVCS, deeply virtual meson production (DVMP) and timelike Compton scattering (TCS).

The formal definition of GPDs in terms of matrix elements can be found in Ref.~\cite{Diehl:2003ny} and we will use the same notations. From the point of view of the current EMT studies, we are interested in the four leading-twist chiral-even GPDs: $H^{a}(x, \xi, t)$, $E^{a}(x, \xi, t)$, $\widetilde{H}^{a}(x, \xi, t)$ and $\widetilde{E}^{a}(x, \xi, t)$ where $a=q, g$. Here, $x$ is the average longitudinal light-front momentum fraction of the active parton and $\xi$ is the skewness variable describing the transfer of longitudinal light-front momentum to the system. GPDs also depend on both factorisation, $\MuF$, and renormalisation, $\MuR$, scales, which will be implicit (unless when specifically needed) to keep concise notations. 

The connection between GPDs and GFFs $A_{a}(t)$, $B_{a}(t)$ and $C_{a}(t)$ is the following \cite{Diehl:2003ny}:
\begin{align}
\label{eq:mel1H}
\int\ud x\,x\,H^q(x,\xi,t)&=A_q(t)+4\xi^2C_q(t) \,, \\
\int\ud x\,x\,E^q(x,\xi,t)&=B_q(t)-4\xi^2C_q(t) \,, \\
\sum_q\int\ud x\,\tilde H^q(x,\xi,t)&=G_A(t) \,,
\end{align}
for quarks, and:
\begin{gather}
\label{eq:mel0H}
\int\ud x\,H^g(x,\xi,t)=A_g(t)+4\xi^2C_g(t) \,, \\
\int\ud x\,E^g(x,\xi,t)=B_g(t)-4\xi^2C_g(t) \,,
\end{gather}
for gluons.
The GPD $\widetilde{E}^{a}(x, \xi, t)$ does not relate to these GFFs\footnote{ $\widetilde{E}^{q}(x, \xi, t)$ is however related to some GFF associated with the EMT for polarised quarks~\cite{Lorce:2014mxa}.}, but is a significant ingredient in the description of DVCS amplitudes and therefore important in the phenomenological applications. 
\newline

We focus now on the extraction of the GFF $C_{a}(t)$ from DVCS data. The DVCS amplitude can be parameterised in terms of structure functions known as Compton form factors (CFFs). For illustration we only give the formulas for the CFF $\mathcal{H}(\xi, t)$ related to the GPD $H^{a}(x, \xi, t)$:  
\begin{gather}
\cH(\xi,t,Q^2) = 
\sum_q \cH^q(\xi,t,Q^2)
+ \cH^g(\xi,t,Q^2) \,,
\end{gather}
where the quark $\cH^q$ and gluon $\cH^g$ contributions read:
\begin{eqnarray}
\cH^q(\xi,t,Q^2) 
& = & 
\int_{-1}^{~1} \frac{\mathrm{d}x}{\xi} \bigg\{
T^q\!\left(\frac{x}{\xi}, \MuF^2, Q^2\right) H^q(x, \xi, t,\MuF^2) \bigg\} \;, \nonumber \\
& & \label{eq:gpd:cff_q_def} \\
\cH^g(\xi,t,Q^2) 
& = & 
\int_{-1}^{~1} \frac{\mathrm{d}x}{\xi} \bigg\{
T^g\!\left(\frac{x}{\xi}, \MuF^2, Q^2\right) \frac{H^g(x, \xi, t,\MuF^2)}{x} \bigg\} \,. \nonumber \\
& & \label{eq:gpd:cff_g_def}
\end{eqnarray}
Here, $Q^2$ is the virtuality of the photon mediating the interaction between the lepton beam and the proton target in DVCS. The functions $T^q$ and $T^g$ are the renormalised coefficient functions (see Ref.~\cite{Moutarde:2013qs} and refs. therein, in slightly different form -- here we added the explicit factors of $x$ and $\xi$ to emphasize the $\frac{x}{\xi}$ dependence of $T^{q,g}$) which for the LO description of the DVCS hard scattering kernel read:
\begin{gather}
T^q = -e_q^2\,\xi /(x+\xi-i\epsilon) - (x \rightarrow -x)\,, \\
T^g = 0 \,,
\end{gather}
where $e_q$ is the quark fractional electric charge in units of the proton charge $|e|$.

In Eqs. \eqref{eq:gpd:cff_q_def} and \eqref{eq:gpd:cff_g_def} we see that CFFs are convolutions of GPDs. The so-called deconvolution problem, which consists in extracting GPDs from CFFs, is far from trivial, and is still an open question today. In the absence of non-parametric extractions of GPDs, an experimental determination of most GFFs thus comes with a significant model uncertainty. However it is possible to access the GFF $C_a(t)$ at the level of amplitudes, \ie without the deconvolution of neither $H^{a}(x, \xi, t)$ nor $E^{a}(x, \xi, t)$. This allows one in principle to avoid any model dependence in the extraction of $C_a(t)$, which makes this situation unique in the context of GPD studies. 

Thanks to the analytic properties of CFFs \cite{Teryaev:2005uj, Anikin:2007yh, Diehl:2007jb}, a dispersion relation connects the real and imaginary parts of the CFF $\cH$, which reads: 
\begin{align}
\label{eq:dr}
 \mathcal{C}_H(t,Q^2)
=  \mathrm{Re}\,\mathcal{H}(\xi, t, Q^2)
\nn\\
- \frac{1}{\pi} \fint_{0}^{1}\mathrm{d}\xi'\,
\mathrm{Im}\,\mathcal{H}(\xi', t, Q^2)
\left(
\frac{1}{\xi-\xi'} -
\frac{1}{\xi+\xi'}
\right) \,,
\end{align}   
where $\fint$ denotes the principal value integral and $\mathcal{C}_H(t,Q^2)$ is the so-called subtraction constant (here constant over $\xi$). Similar equations hold for the CFFs $\cE$, $\mathcal{\widetilde{H}}$ and $\mathcal{\widetilde{E}}$, but with the opposite subtraction constant for $\mathcal{E}$ and null subtraction constants for $\mathcal{\widetilde{H}}$ and $\mathcal{\widetilde{E}}$.

The $D$-term is defined by the coefficients proportional to the highest powers of $\xi$ in the polynomiality relations of a given GPD \cite{Diehl:2003ny}. It therefore allows one to access the GFF $C_a(t)$ via Eqs. \eqref{eq:mel1H} and \eqref{eq:mel0H}. Remarkably the subtraction constant $\mathcal{C}_H$ is related (to any order in QCD perturbation theory) to the quark and gluon $D$-terms in the following way:
\begin{equation}
\label{eq:quark-gluon-subtraction-constants}
   \mathcal{C}_H(t,Q^2) = \sum_q \mathcal{C}^q_H(t,Q^2) + \mathcal{C}^g_H(t,Q^2) \;,
\end{equation}
with (omitting the dependence on $\MuF^2$ and $Q^2$ for the sake of conciseness):
\begin{eqnarray}
\mathcal{C}^q_H(t,Q^2)
& = & 
\frac{2}{\pi} \int_1^\infty \mathrm{d}\omega \, \mathrm{Im} T^q(\omega) \int_{-1}^{~1}\mathrm{d}z \, \frac{D^q_{\mathrm{term}}(z,t)}{\omega-z} \;, \nonumber \\
& & \\
\mathcal{C}^g_H(t,Q^2)
& = & 
\frac{2}{\pi} \int_1^\infty \frac{\mathrm{d}\omega}{\omega} \, \mathrm{Im} T^g(\omega) \int_{-1}^{~1}\mathrm{d}z \, \frac{D^g_{\mathrm{term}}(z,t)}{\omega-z} \;. \nonumber \\
& & 
\end{eqnarray}
At LO the explicit gluon dependence of the subtraction constant vanishes and the imaginary part of the quark coefficient function merely selects the value $\omega = 1$:
\begin{gather}
\label{eq:SC_D}
\mathcal{C}_H(t,Q^2)
= 2\sum_q e_q^2\,\int_{-1}^{~1}\ud z\,\frac{D_{\mathrm{term}}^q(z,t,\MuF^2 \equiv Q^2)}{1-z} \nonumber \,.
\end{gather}
Here, $Q^{2}$ is identified with $\MuF^2$. Because Gegenbauer polynomials diagonalise the ERBL evolution equations at LO, the $D$-term is often expanded into a series of these polynomials \cite{Goeke:2001tz}:
\begin{align}
\label{eq:Dq}
D_{\mathrm{term}}^q(z,t,\MuF^2)
& = (1-z^2) \sum_{\text{odd } n}d_n^q(t,\MuF^2)\, C_n^{3/2}(z) \,,
\\
\label{eq:Dg}
D_{\mathrm{term}}^g(z,t,\MuF^2)
&= \frac{3}{2} (1-z^2)^2 \sum_{\text{odd } n}d_n^g(t,\MuF^2)\, C_{n-1}^{5/2}(z) \,,
\end{align}
where $d_n^q(t,\MuF^2)$, $d_n^g(t,\MuF^2)$ are the coefficients of the expansion. The values of the first three flavour-singlet coefficients:
\begin{eqnarray}
d_1^{u+d}(0 , \mu_0^2) & \simeq & - 4.0, \label{eq:ChQSM-d1-low-scale} \\
d_3^{u+d}(0, \mu_0^2) & \simeq & - 1.2, \label{eq:ChQSM-d3-low-scale} \\
d_5^{u+d}(0, \mu_0^2) & \simeq & - 0.4. \label{eq:ChQSM-d5-low-scale} 
\end{eqnarray}
from the chiral quark-soliton model ($\chi$QSM) at a low scale $\mu_0 \simeq 600~\mathrm{MeV}$, are reported in Ref.~\cite{Goeke:2001tz}. We observe that each coefficient is roughly three times bigger than its successor.

Gegenbauer polynomials $C_n^{\alpha}(z)$ are classically defined as a family of orthogonal polynomials with respect to the weight function $(1-z^2)^{\alpha-1/2}$, but they can also be regarded as coefficients of the formal series (in the variable $t$) of the following generating function:
\begin{equation}
    \frac{1}{(1 - 2 z t + t^2)^\alpha} = \sum_{n=0}^\infty C_n^{\alpha}(z) t^n \;.
\end{equation}
For $|t|<1$ we observe that:
\begin{align}
    &\sum_{n=0}^\infty \int_{-1}^1 \ud z\,\frac{(1-z^2)C_n^{3/2}(z)}{1-z}\,t^n\nonumber\\
    &=\int_{-1}^{+1} \mathrm{d}z \, \frac{1+z}{(1 - 2 z t + t^2)^{3/2}} = \frac{2}{1-t} = \sum_{n=0}^\infty 2t^n \;.
\end{align}
Therefore the DVCS subtraction constant at LO can be conveniently written as:
\begin{gather}
    \mathcal{C}_H(t,Q^2) = 4\sum_q e_q^2\,\sum_{\text{odd } n} d_n^q(t,\MuF^2 \equiv Q^2) \,.
    \label{eq:SC_d}
\end{gather}
The first element of this expansion carries information about the GFF $C_{a}(t)$. By definition of the $D$-term, using (\ref{eq:mel1H}) and (\ref{eq:mel0H}):
\begin{align}
\int_{-1}^1 \mathrm{d}z\,  z D_{\mathrm{term}}^q(z,t) = 4 C_q(t) \,, \\
\int_{-1}^1 \mathrm{d}z\,  D_{\mathrm{term}}^g(z,t) = 4 C_g(t) \,. 
\end{align}
Since $C_1^{3/2}(z) = 3z$ and $C_0^{5/2}(z) = 1$, the definition of Gegenbauer polynomials $C_n^{3/2}(z)$ and $C_n^{5/2}(z)$ as orthogonal polynomials associated to the respective weights $1-z^2$ and $(1-z^2)^2$, complemented by the expansions (\ref{eq:Dq}) and (\ref{eq:Dg}), yields:
\begin{gather}
d^q_1(t,\MuF^2) = 5C_q(t,\MuF^2) \,, \label{eq:d1q-to-quark-GFF} \\
d^g_1(t,\MuF^2) = 5C_g(t,\MuF^2) \,. \label{eq:d1g-to-gluon-GFF}
\end{gather}
Note that:
\begin{eqnarray}
d(t) & = & \sum_{q}d_1^{q}(t, \MuF^2) + d_1^{g}(t, \MuF^2)  \label{eq:def-d-total} \\
& = & 5 \Big( \sum_{q}C_q(t, \MuF^2) + C_g(t, \MuF^2) \Big) \nonumber \;,
\end{eqnarray}
becomes scale independent as it is one of the gravitational form factors of the full EMT. 

Through Eq.~\eqref{eq:sbis}, the two relations \eqref{eq:d1q-to-quark-GFF} and \eqref{eq:d1g-to-gluon-GFF} provide a direct handle on the distribution of pressure anisotropy (for quarks and gluons) in the proton:
\begin{eqnarray}
s_q(r)
& = & -\frac{4 M}{5 r^2}\int\frac{\ud^3\boldsymbol\Delta}{(2\pi)^3}\,
e^{-i\boldsymbol\Delta\cdot\boldsymbol r}\,\frac{t^{-1/2}}{M^2}\frac{\ud^2}{\ud t^2}\!\left[t^{5/2}\,d^q_1(t)\right] \;, \nonumber \\
& & 
\label{eq:pressure_profile_q}\\
s_g(r)
& = & -\frac{4 M}{5 r^2}\int\frac{\ud^3\boldsymbol\Delta}{(2\pi)^3}\,
e^{-i\boldsymbol\Delta\cdot\boldsymbol r}\,\frac{t^{-1/2}}{M^2}\frac{\ud^2}{\ud t^2}\!\left[t^{5/2}\,d^g_1(t)\right] \;, \nonumber \\
& & 
\label{eq:pressure_profile_g}
\end{eqnarray}
where the dependence on the factorisation scale $\MuF^2$ is implicit. The phenomenological challenge of the experimental determination of pressure forces in the proton thus boils down to the extraction of the first term in the expansion \eqref{eq:SC_d} of the (measurable) DVCS subtraction constant. In the next section we will show that each term in this expansion behaves differently under LO evolution, allowing one to separate $d^q_1(t,\MuF^2)$ and $d^g_1(t,\MuF^2)$ from the rest of the series \eqref{eq:SC_d}.

%
%
\section{Evolution of $D$-term and DVCS subtraction constant}
\label{sec:evolution}

The evolution of the $D$-term (and therefore of the DVCS subtraction constant) is governed by the ERBL evolution equations~\cite{Efremov:1979qk, Lepage:1979zb}. These equations can be explicitly solved at LO once the $D$-term is expressed in terms of Gegenbauer polynomials, just like in Eqs.~\eqref{eq:Dq} and \eqref{eq:Dg}. We follow the presentation of Ref.~\cite{Diehl:2003ny}.

Let $n_f$ denote the number of active quark flavours. The anomalous dimensions $\gamma_{n}$, $\gamma_n^{\pm}$ driving the LO evolution read:
\begin{align}
    \gamma_{n} =&\, \gamma_{QQ}(n) \,,
    \\
    \gamma_n^{\pm} =&\,
    \frac{1}{2}\bigg(
    \gamma_{QQ}(n) + \gamma_{GG}(n)
    \nn\\
    &\pm \sqrt{[\gamma_{QQ}(n) - \gamma_{GG}(n)]^2 + 4\gamma_{QG}(n)\gamma_{GQ}(n)}
    \,\bigg) \,,
\end{align}
where \cite{Diehl:2000uv}:
\begin{align}
\gamma_{QQ}(n)
&= C_{F}\left( \frac{1}{2} - \frac{1}{(n+1)(n+2)} + 2\sum_{k=2}^{n+1} \frac{1}{k}\right) \,,
\\
\gamma_{QG}(n)
&= -n_{f}\, T_{F}\frac{n^2+3n+4}{n(n+1)(n+2)} \,,
\\
\gamma_{GQ}(n) &= -2\, C_{F} \frac{n^2+3n+4}{(n+1)(n+2)(n+3)} \,,
\\
\gamma_{GG}(n) &= \frac{2}{3} n_{f} T_{F} +C_{A}
\nn\\
    &\times
    \left(
    \frac{1}{6} -
    \frac{2}{n(n+1)} -
    \frac{2}{(n+2)(n+3)} +
    2 \sum_{k=2}^{n+1} \frac{1}{k} 
    \right) \,.
\end{align}
Here $C_{F} = \nicefrac{4}{3}$, $T_{F} = \nicefrac{1}{2}$ and $C_{A} = 3$. 

At the leading logarithmic accuracy and for a non-singlet combination of $q_{1}$ and $q_{2}$ quark flavours, the evolution from the initial scale $\MuFRef^2$ to the final scale $\MuF^2$ is given by:
\begin{align}
d_n^{q_1}&(t, \MuF^2) - d_n^{q_2}(t, \MuF^2)
\nn\\
&= 
    \left[
    d_n^{q_1}(t, \MuFRef^2) - d_n^{q_2}(t, \MuFRef^2)
    \right]
    \left(
    \frac{
    \alphas(\MuF^2)}{\alphas(\MuFRef^2)}
    \right)^{\textstyle\frac{2\gamma_n}{\beta_0}} \,.
    \label{eq:non-singlet-evolution}
\end{align}
Introducing the mixing coefficients, $a_n^{\pm}$:
\begin{gather}
    a_n^\pm =
    2\,\frac{n_f}{n}\,\frac{\gamma_n^{\pm}-\gamma_{n}}{\gamma_{QG}(n)} \,,
\end{gather}
the singlet combination of coefficients of the Gegenbauer expansion \eqref{eq:Dq}, and the gluon coefficients of the analogous expansion \eqref{eq:Dg}, read:
\begin{gather}
    \frac{1}{n_f} \sum_q d_n^q (t, \MuF^2) = d_n^{+}(t,\MuF^2) +  d_n^{-}(t,\MuF^2) \,, \label{eq:d-singlet}
    \\
    d_n^g(t,\MuF^2) = a_n^{+} d_n^{+}(t,\MuF^2) + a_n^{-} d_n^{-}(t,\MuF^2) \,. \label{eq:d-gluon}
\end{gather}
This decomposition is convenient since the coefficients $d_n^\pm$ evolve multiplicatively:
\begin{gather}
    d_n^{\pm}(t,\MuF^2) = d_n^{\pm}(t,\MuFRef^2) \left(\frac{\alphas(\MuF^2)}{\alpha_s(\MuFRef^2)}\right)^{\textstyle\frac{2\gamma_n^{\pm}}{\beta_0}} \,.
\end{gather}
The inversion of the system Eqs.~\eqref{eq:d-singlet}-\eqref{eq:d-gluon} is straightforward:
\begin{gather}
    d_n^\pm (t, \MuFRef^2)
    = \pm \frac{d_n^g(t, \MuFRef^2) -  \displaystyle \frac{a_n^\mp}{n_f} \displaystyle \sum_q d_n^q(t,\MuFRef^2)}{a_n^+ - a_n^-} \,.
\end{gather}

It is useful to look at the behavior of the $D$-term in the $\MuF^2 \to \infty$ limit. Since all the anomalous dimensions $\gamma_n^{\pm}$ are positive, except $\gamma_{1}^{-} = 0$, all $d_{n}^{q}$ and $d_{n}^{g}$ coefficients vanish in this limit, except maybe the following two~\cite{Goeke:2001tz}:
\begin{align}
\frac{1}{n_{f}}\sum_{q}d_1^q(t, \MuF^2) 
&\xrightarrow{\MuF^2 \to \infty} \, d_1^{-}(t, \MuFRef^2) \,,
\\
d_1^g(t, \MuF^2)
&\xrightarrow{\MuF^2 \to \infty} \, a_1^{-} d_1^{-}(t, \MuFRef^2) \,,
\end{align}
with:
\begin{align}
d_1^{-}(t, \MuFRef^2)
&= d(t, \MuFRef^2) \frac{T_F}{n_{f}\,T_F + 2 C_{F}} \,, 
\\
a_1^{-} d_1^{-}(t, \MuFRef^2)
&= d(t, \MuFRef^2) \frac{2 C_F}{n_{f}\,T_F + 2 C_{F}} \,,
\end{align}
where $d(t, \MuFRef^2)$ has been defined in Eq.~\eqref{eq:def-d-total}.

We conclude with the following remark. If the light flavours contribute equally, \ie if  $d_n^{q_1}(t, \MuFRef^2)$ = $d_n^{q_2}(t, \MuFRef^2)$ for the two light flavours $q_1$ and $q_2$, then the non-singlet combination \eqref{eq:non-singlet-evolution} identically vanishes and the left-hand side of the singlet combination (\ref{eq:d-singlet}) reduces to the common value for the considered light flavours.

%
%

\section{Modelling of DVCS subtraction constant}
\label{sec:model}

We have seen that an unbiased extraction at LO is theoretically possible through the lever arm in $Q^2$ of the subtraction constant, but the weak $t$ and $Q^2$ dependence of current experimental data motivates us to introduce a simplified modelling of the $D$-term. Those simplifications are typical of analyses like the present one. We list below these assumptions and discuss their importance on the extraction of $D$-term information.
\begin{enumerate}
  \item  We restrict the analysis only to the first element of the Gegenbauer expansions (\ref{eq:Dq}) and (\ref{eq:Dg}), \ie only $d_{1}^{q}(t, \MuF^2)$ is directly fitted. 
  
  We may hope that these Gegenbauer expansions converge, and consequently that their general terms asymptotically vanish. However this does not justify a strict dominance of the first terms of the series; extrapolating the successive ratios ($\simeq 1/3$) of chiral quark-soliton model estimates (\ref{eq:ChQSM-d1-low-scale})-(\ref{eq:ChQSM-d5-low-scale}) to infinite orders, we may think that retaining only $d_1$ generates a 50\% systematic uncertainty. In order to fit coefficients proportional to other terms of these expansions, in particular $d_{3}^{q}(t, \MuF^2)$, one needs extra constraints separating various contributions, in this case either the $t$ or $\MuF^2$ unique dependence. Since not much is known about the $t$-dependence from first principles (in particular there is no reason to think that all coefficients obey the same $t$-dependence), GPD evolution provides a natural tool to extract $d_{3}^{q}(t, \MuF^2)$ and even higher order terms. To achieve this, one needs both precise data and a large arm in $Q^2$ to study the evolution carefully. In the following we show that the current data does not allow for such kind of analysis. EIC and EIcC are natural candidates for changing this conclusion, but the quantitative assessment of their impact goes beyond the scope of the present study.
  
  \item  We assume an equal contribution $d_{n}^{uds}(t, \MuF^2)$ of light quarks to each coefficient $d_{n}^{q}$ of the $D$-term expansion \eqref{eq:Dq}, \ie:
  \begin{align}
    d_{n}^{u} = d_{n}^{d} = d_{n}^{s} \equiv d_{n}^{uds} \,.
  \end{align}
  In particular, the previous assumption focuses our study on $d_{1}^{uds}(t, \MuF^2)$ satisfying:
  \begin{align}
    d_{1}^{u} = d_{1}^{d} = d_{1}^{s} \equiv d_{1}^{uds} \,.
  \end{align}

  This is a common assumption, met \eg in the chiral quark–soliton model~\cite{Petrov:1998kf, Goeke:2001tz}. We stress that disentangling the separate quark flavour contributions is not possible at LO with DVCS alone, but should be feasible for instance by studying hard exclusive meson production data. Early studies of this type~\cite{Teryaev:2013dka} demonstrate the potential of such kind of analysis. 
  
  In such a case, the relation \eqref{eq:SC_d} between the subtraction constant and the coefficients $d_1^q$ becomes:
\begin{equation}
    \mathcal{C}_H(t,Q^2) = \frac{8}{3} d_1^{uds}(t, Q^2) = \frac{8}{9} \sum_{q} d_{1}^{q}(t, Q^2) \;. \label{eq:subtraction-vs-d1-uds}
\end{equation}
  As we will see in Sec.~\ref{sec:results}, some studies consider only two active quark flavours (\ie no strange quark), which still equally contribute to the $D$-term. With an obvious adaptation of notations, the relation~\eqref{eq:SC_d}  now reads:
\begin{equation}
    \mathcal{C}_H(t,Q^2) = \frac{20}{9} d_1^{ud}(t, Q^2) = \frac{10}{9} \sum_{q} d_{1}^{q}(t, Q^2) \;. \label{eq:subtraction-vs-d1-ud} 
\end{equation}
  
  \item Because of the absence of a direct sensitivity to the gluon $D$-term in a LO analysis of DVCS, $d_{1}^{g}$ is not fitted to the data. 
  
  Instead, $d_{1}^{g}$ is  radiatively generated starting from a low factorisation scale where a valence quark picture of the proton should hold. Although it contradicts the conclusion of Ref.~\cite{Diehl:2019fsz}, this assumption is frequently met in the computation of various parton distribution functions from quark models, and does not prevent an analysis of the existing DVCS data. Here we set: 
   \begin{equation}
   d_{1}^{g}(t, \MuFRef^{2}) = 0 \textrm{ at } \MuFRef^{2} = 0.1~\mathrm{GeV}^2\,.
   \end{equation}
   The sensitivity of this choice on the extraction of $d_{1}^{uds}(t, \MuF^2)$ is discussed in the following. The radiative generation of gluons will manifest itself with a non-vanishing $d_{1}^{g}(t, \MuF^{2})$ at scales $\MuF^2 > \MuFRef^{2}$.
   
  \item Since a significant amount of current DVCS data possess $Q^{2}$ values larger than the squared mass of the charm quark, $m_{c}^{2} = (1.28~\pm~0.03)^2~\mathrm{GeV}^{2}$, $d_{1}^{c}(t, \MuF^2)$ may contribute to the subtraction constant. 
  
  This contribution is expected to be negligible in the considered kinematic range. However, it can be conveniently generated by the evolution equations, similarly to gluons. In this analysis we explore this possibility with the following boundary condition:
  \begin{equation}
  d_{1}^{c}(t, m_{c}^{2}) = 0 \,.
  \end{equation}  
  The radiative generation of charm quarks will manifest itself with a non-vanishing $d_{1}^{c}(t, \MuF^{2})$ at scales $\MuF^2 > m_c^{2}$.
  
  \item We adopt a multipole form for the $t$-dependence of $d(t, \MuF^2)$ for $d \in \{d_{1}^{uds}, d_{3}^{uds}, d_1^{c}, d_1^{g}\}$:
  \begin{equation}
  \label{eq:tripol}
  d(t, \MuF^2) = d(\MuF^2)
  \left( 1 - \displaystyle\frac{t}{\Lambda^{2}} \right)^{-\alpha}\,,
  \end{equation}
  where $d_{1}^{uds}(\MuF^2)$, $d_{3}^{uds}(\MuF^2)$, $d_1^{c}(\MuF^2)$ and $d_1^{g}(\MuF^2)$ are parameters to be determined. Unless explicitly stated otherwise, the parameters $\Lambda = 0.8~\mathrm{GeV}$ and $\alpha = 3$ are kept fixed in our fits. In the following we will refer to it as the tripole Ansatz. The value of $\Lambda$ is motivated by the chiral quark-soliton model \cite{Goeke:2007fp, Polyakov:2018exb}, while that for $\alpha$ ensures a realistic shape of the pressure distribution at large $t$~\cite{Lorce:2018egm}. 
\end{enumerate}

We will also explore the possibility of extracting one of the parameters $d_1^{g}(\MuF^2)$, $d_{3}^{uds}(\MuF^2)$, $\Lambda$ or $\alpha$ together with $d_{1}^{uds}(\MuF^2)$ in the following.

%
%
\section{Input from global fits to DVCS data}
\label{sec:fits}

The procedure described in this section is used to find replicas representing the subtraction constant from a given set of replicas representing CFFs. In particular we detail the implementation of the multipole Ansatz \eqref{eq:tripol} driving the kinematic extrapolation to vanishing $t$. 

In Ref.~\cite{Moutarde:2019tqa}, 30 distinct DVCS observables spread over 2,500 kinematic configurations and collected over 17 years were jointly analysed in terms of CFFs relying on a neural network approach. In this study the real and imaginary parts of each CFF were independently described and simultaneously fitted to experimental data. All four leading-twist and chiral-even GPDs were considered, and no assumption was made beyond the validity of a leading-twist analysis. The uncertainties of experimental data are reflected through a set of 101 replicas for each of the eight extracted functions (\ie for real and imaginary parts of each CFF).

Let us denote the replicas associated to the real and imaginary parts of the CFF $\mathcal{H}$ by:
\begin{align}
\mathrm{Re}\,\mathcal{H}_{i}^{\mathrm{NN}}(\xi, t, Q^2)
\quad\textrm{and}\quad
\mathrm{Im}\,\mathcal{H}_{i}^{\mathrm{NN}}(\xi, t, Q^2)\,,
\end{align}
where $i=0, \ldots, 100$, and where the superscript $\mathrm{NN}$ indicates that a given quantity is obtained in the neural network analysis. Here, each replica is a function of the three variables $\xi$, $t$ and $Q^2$, and represents a single neural network built of i) three input neurons receiving the values of $\xi$, $t$ and $Q^{2}$, ii) one hidden layer with six neurons, and iii) one output neuron returning either the real or imaginary part of the CFF $\mathcal{H}$. The parameters of this function, which are weights and biases ``trained'' to the experimental data, are real numbers  (for details see Ref. \cite{Moutarde:2019tqa}). Those numbers are not intuitive in the sense that they do not carry any physical meaning -- they are not ``human readable'' or interpretable. The replica for $i = 0$ is obtained from a direct fit to the experimental data, while those for $i \neq 0$ are obtained from data where the central values are randomly smeared according to uncertainties. In this case $i$ is used to distinguish between the unique values of the random seed used in the smearing procedure.  

The subtraction constant is evaluated from a pair of $i$-th replicas ($0 \leq i \leq 100$) using the dispersion relation: 
\begin{align}
\label{eq:dr_used}
 \mathcal{C}_{H,i}(\xi,t,Q^2)(t, Q^2)
= \mathrm{Re}\,\mathcal{H}_{i}^{\mathrm{NN}}(\xi, t, Q^2)
\nn\\
- \frac{1}{\pi} \fint_{\epsilon}^{1}\mathrm{d}\xi'\,
\mathrm{Im}\,\mathcal{H}_{i}^{\mathrm{NN}}(\xi, t, Q^2)
\left(
\frac{1}{\xi-\xi'} -
\frac{1}{\xi+\xi'}
\right) \,,
\end{align}
where the $\xi'$-integration ranges between $\epsilon = 10^{-6}$ and 1 instead of 0 and 1 as in Eq.~\eqref{eq:dr}. This corresponds to the kinematic domain probed by the replicas in the global CFF fit to DVCS data. This replacement of the integration range introduces a negligible $\mathcal{O}(1\%)$ bias, see Ref.~\cite{Moutarde:2019tqa} for further details. We remind that dispersion relations derive from first principles, which makes this approach model-independent. 

In this analysis, the expectation value and the variance of the subtraction constant distribution at a kinematic configuration $(\xi, t, Q^2)$ are estimated with the empirical mean and the standard deviation of a replica subset of $\big(\mathrm{Re}\mathcal{H}_{i}^{\mathrm{NN}}, \mathrm{Im}\mathcal{H}_{i}^{\mathrm{NN}}\big)_{0 \leq i \leq 100}$ at the same kinematic configuration:
\begin{align}
\label{eq:mean}
&\mu_{\mathcal{C}_{H}}(\xi, t, Q^2) = \frac{1}{|I|}\sum_{i \in I} \mathcal{C}_{H,i}(\xi, t, Q^2) \,, 
\\
\label{eq:sigma}
&\sigma_{\mathcal{C}_{H}}(\xi, t, Q^2)
\nn\\
&\quad= \sqrt{\frac{1}{|I|}\sum_{i \in I} \left[\mathcal{C}_{H,i}(\xi, t, Q^2) - \mu_{\mathcal{C}_{H}}(\xi, t, Q^2)\right]^2} \,,
\end{align}
where $|I| \leq 101$ is the number of replicas $i$ in the subset $I$ of $\{0, 1, \ldots, 100\}$ used in the estimation. Using most of, but not all, the available replicas is motivated by the presence of outliers: it happens sometimes that some replicas give exceptional or ``exotic'' values widely separated from those returned by the rest of the replica population. This typically signals problems with the supervised training of neural networks, either because some kinematic regions are not sufficiently covered by data, or \eg when the involved training algorithm exceptionally converges to a local minimum \cite{Moutarde:2019tqa}. The procedure to remove the outliers is described in the next paragraph.

It is assumed that at a given $(\xi, t, Q^2)$-point the replicas return normally distributed values, which in particular allow for a straightforward assignment of $68\%$ confidence level to the range of $\mu_{\mathcal{C}_{H}}(\xi, t, Q^2) \pm \sigma_{\mathcal{C}_{H}}(\xi, t, Q^2)$, and a simple propagation of uncertainties since the probability distribution of replicas becomes unambiguously characterised by the two parameters $\mu_{\mathcal{C}_{H}}(\xi, t, Q^2)$ and $\sigma_{\mathcal{C}_{H}}(\xi, t, Q^2)$. This assumption is in general fulfilled, however not in all cases, because of the presence of outliers. Since these issues are in general local in the $3\mathrm{D}$ phase-space of kinematic configurations $(\xi, t, Q^2)$, it is difficult to identify problematic replicas and entirely remove them from the analysis. Such procedure could also introduce a bias. Instead, the outliers are locally detected and removed with the classical three-sigma rule. This rule is known from big data analytics \cite{IlyasC19}, where it is successfully used to improve the quality of data under consideration. The procedure is iterative. Starting from $I = \{0, 1, \ldots, 100\}$ (hence $|I| = 101$), it consists of the following steps:
\begin{enumerate}[labelwidth=\widthof{$3\mathrm{a.}$},leftmargin=!]
  \item[$1.$] evaluate the mean $\mu$, and standard deviation $\sigma$, of a given sample $\big(\mathcal{C}_{H,i}(\xi,t,Q^2)\big)_{0 \leq i \leq 100}$ using Eqs.~(\ref{eq:mean}) and (\ref{eq:sigma}),
  \item[$2.$] remove from that sample all replicas $i$ for which $\mathcal{C}_{H,i}(\xi,t,Q^2)$ does not lay in the $(\mu - 3\sigma, \mu + 3\sigma)$ range,
  \item[$3\mathrm{a.}$] if no elements have been removed, stop the procedure -- the sample is free of outliers -- and retrieve the subset I, the mean $\mu_{\mathcal{C}_{H}}(\xi, t, Q^2)$ and the standard deviation $\sigma_{\mathcal{C}_{H}}(\xi, t, Q^2)$,
  \item[$3\mathrm{b.}$] if any element has been removed, go to $1.$ and proceed with the next steps.
\end{enumerate} 
To some extent, the procedure is equivalent to a fit to a Gaussian Ansatz and a removal of all elements populating the exterior of the $\mu \pm 3\sigma$ range.

We will now fit the multipole Ansatz, see Eq.~\eqref{eq:tripol}, to the subtraction constant replicas obtained in the neural network analysis. 
We work with the kinematic configuration $(\xi_{j}, t_{j}, Q^{2}_{j})_{1 \leq j \leq N_{\mathrm{pts}}}$ where DVCS measurements have been published, with $\xi = x_{\mathrm{Bj}}/(2-x_{\mathrm{Bj}})$ in the Bjorken regime and $N_{\mathrm{pts}} = 277$. We will explore various fitting scenarios where the extracted parameters constitute a subset $Y$ of $\{d_{1}^{uds}(\MuF^2), d_{3}^{uds}(\MuF^2), d_{1}^{g}(\MuF^2), \Lambda, \alpha\}$ for some hadronic scale $\MuF^2 > m_c^{2}$. For each subtraction constant replica $\mathcal{C}_{H,i}$ with $i \in \{0, 1, \ldots, 100\}$, we define the $\chi^{2}$-function $\chi_{i}^{2}(Y)$ for the generated points in the following way:
\begin{gather}
\label{eq:chi2}
\chi_{i}^{2}(Y)
\nn\\
= \sum_{j = 1}^{N_{\mathrm{pts}}}
    \left(
    \frac{
    \mathcal{C}_{H,i}(\xi_{j}, t_{j}, Q^{2}_{j})
    -
    \mathcal{C}_{H, i}^{\mathrm{\alpha P}}(\xi_{j}, t_{j}, Q^{2}_{j}, Y)
    }
    {\sigma_{\mathcal{C}_{H}}(\xi_{j}, t_{j}, Q^{2}_{j})}
    \right)^{2} \,,
\end{gather}
where $(\xi_{j}, t_{j}, Q^{2}_{j})$ is the kinematic configuration generated for the point $j$, $\mathcal{C}_{H,i}(\xi_{j}, t_{j}, Q^{2}_{j})$ and $\sigma_{\mathcal{C}_{H}}(\xi_{j}, t_{j}, Q^{2}_{j})$ are, respectively, the neural network analysis values of the subtraction constant returned by replica $i$ and its uncertainty at point $j$, and $\mathcal{C}_{H, i}^{\mathrm{\alpha P}}(\xi_{j}, t_{j}, Q^{2}_{j}, Y)$ is the value of the subtraction constant from the multipole Ansatz also at point $j$. The $\chi^{2}$-function $\chi_{i}^{2}(Y)$ allows one to constrain the free parameters $Y$ of the tripole or multipole Ansatz. In contrast to the neural network analysis, those parameters possess a physical interpretation, and in particular straightforwardly provide the value of the subtraction constant at $t=0$.

%
%
\section{Results}
\label{sec:results}

\subsection{Nominal fitting scenario}
\label{sec:nominal-fit}

In this analysis, we first consider a nominal fitting scenario where only one parameter is directly fitted to data parameterised with neural networks, namely $d_{1}^{uds}(\MuF^{2})$. The values for gluons, $d_{1}^{g}(\MuF^{2})$, and charm quarks, $d_{1}^{c}(\MuF^{2})$, are indirectly fitted, \ie they are generated ``radiatively'' by the LO evolution using the following boundary conditions:
\begin{flalign}
&d_{1}^{g}(\mu_{\mathrm{F}, 0}^{2}) = 0\,, \\ 
&d_{1}^{c}(m_{c}^{2}) = 0 \,,
\end{flalign}
where $\mu_{\mathrm{F}, 0}^{2} = 0.1~\mathrm{GeV}^{2}$ is the initial factorisation scale squared and where $m_{c}^{2} \simeq 1.64~\mathrm{GeV}^{2}$ is the mass of charm quark squared. For details see Sect.~\ref{sec:model} above. The obtained numerical values are summarised in Tab.~\ref{tab:result:def}, where the values of $d_{1}^{uds}$, $d_{1}^{g}$ and $d_{1}^{c}$ are quoted at the typical hadronic scale $\MuF^2 = 2~\mathrm{GeV}^2$.

\begin{table}[!ht]
\caption{Values of $d_{1}^{uds}$, $d_{1}^c$ and $d_{1}^g$ for light, charm quarks and gluons quoted at $\MuF^2 = 2~\mathrm{GeV}^2$. Those values are obtained with a tripole Ansatz, where only $d_{1}^{uds}(\MuF^2)$ is directly fitted to experimental data parameterised with neural networks.}
\label{tab:result:def}
\begin{center}
\begin{tabular}{@{}cc@{}}
\toprule
\multicolumn{1}{l}{Parameter} & Value \\ \midrule
$d_{1}^{uds}(\MuF^2)$ & $-0.5 \pm 1.2$ \\
$d_{1}^{c}(\MuF^2)$ & $-0.0020 \pm 0.0053$ \\
$d_{1}^{g}(\MuF^2)$ & $-0.6 \pm 1.6$ \\ \bottomrule
\end{tabular}
\end{center}
\end{table}

\begin{figure}[!ht]
\begin{center}
\includegraphics[width=\figWidth]{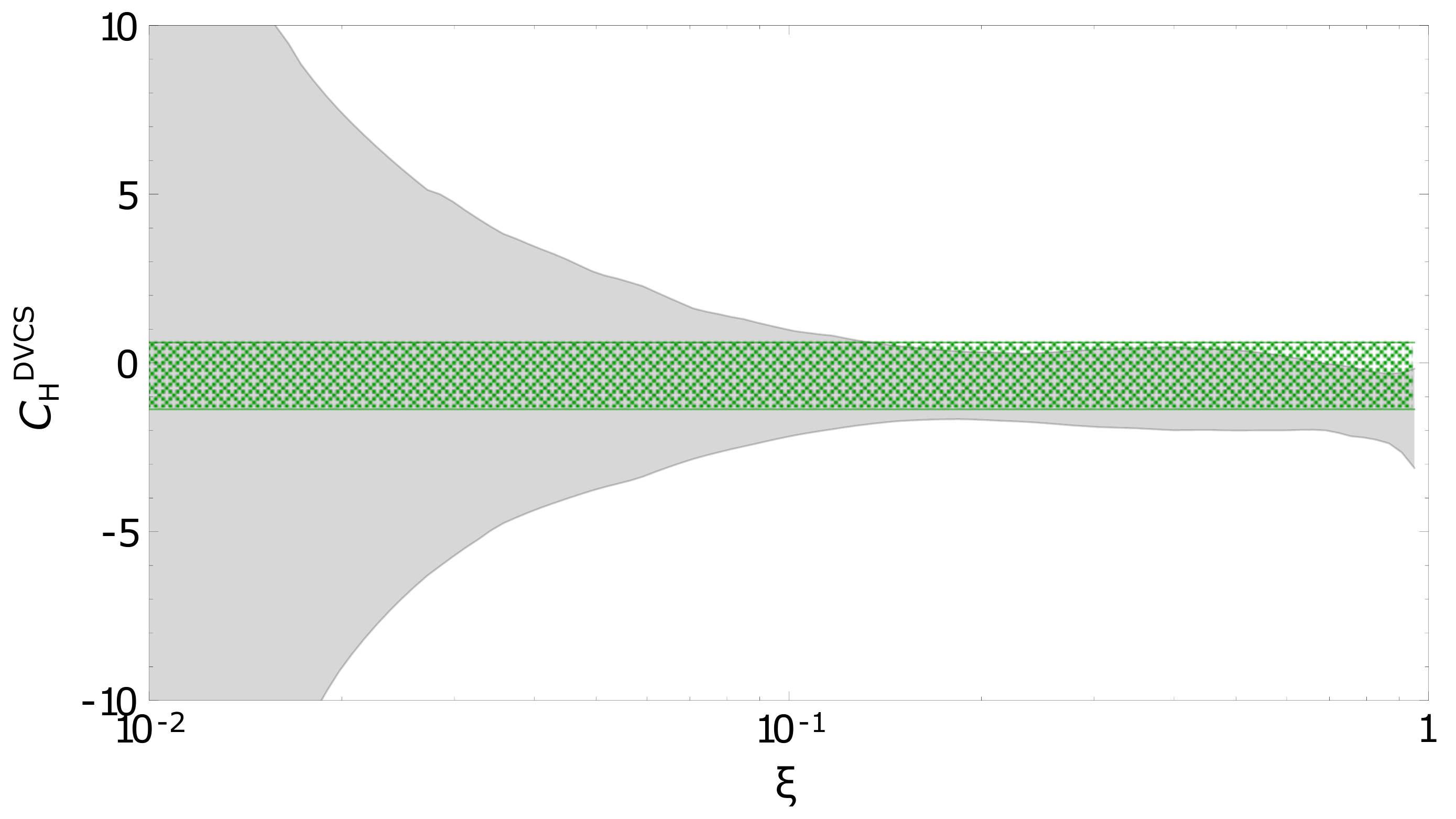}
\includegraphics[width=\figWidth]{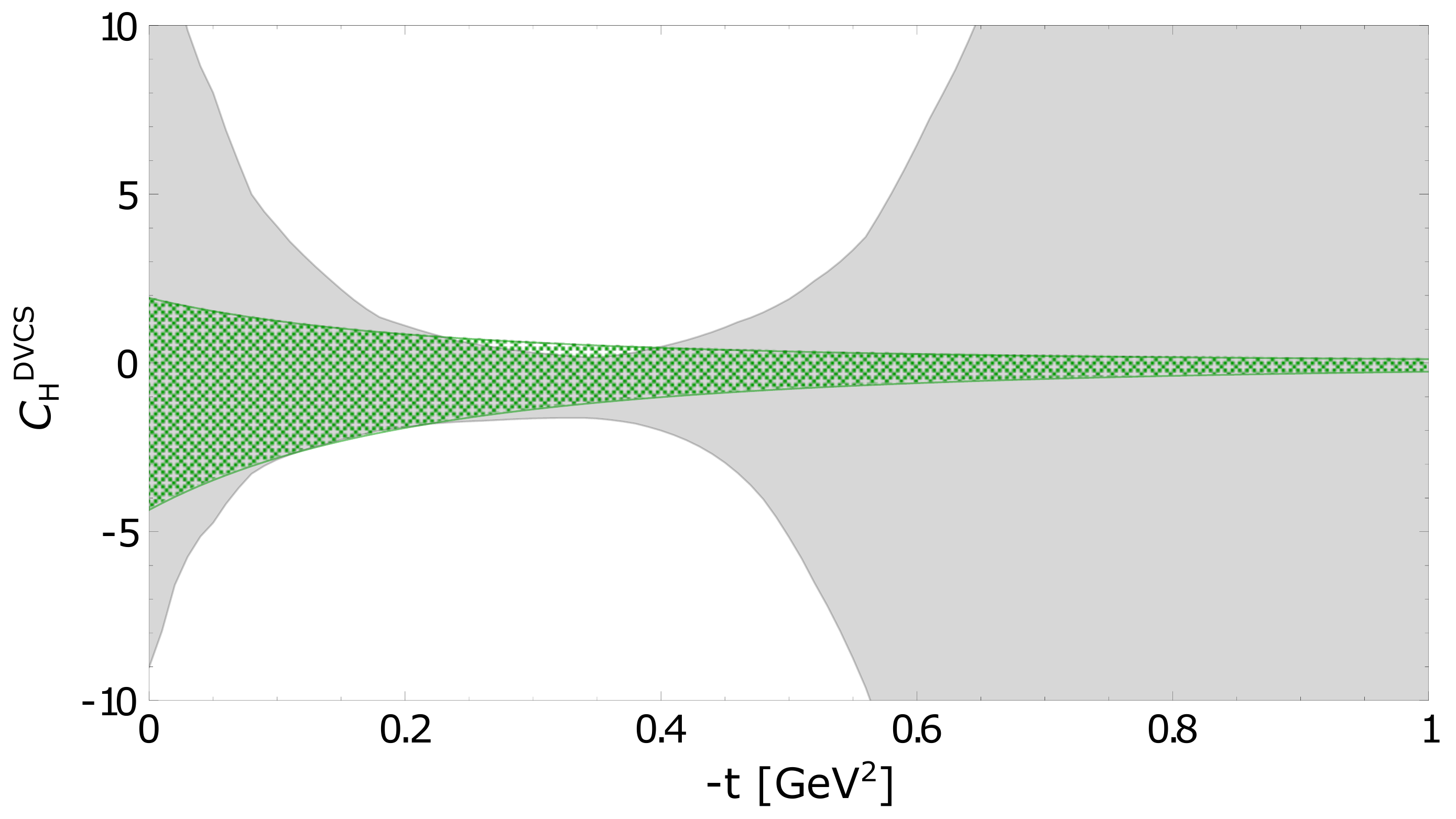}
\includegraphics[width=\figWidth]{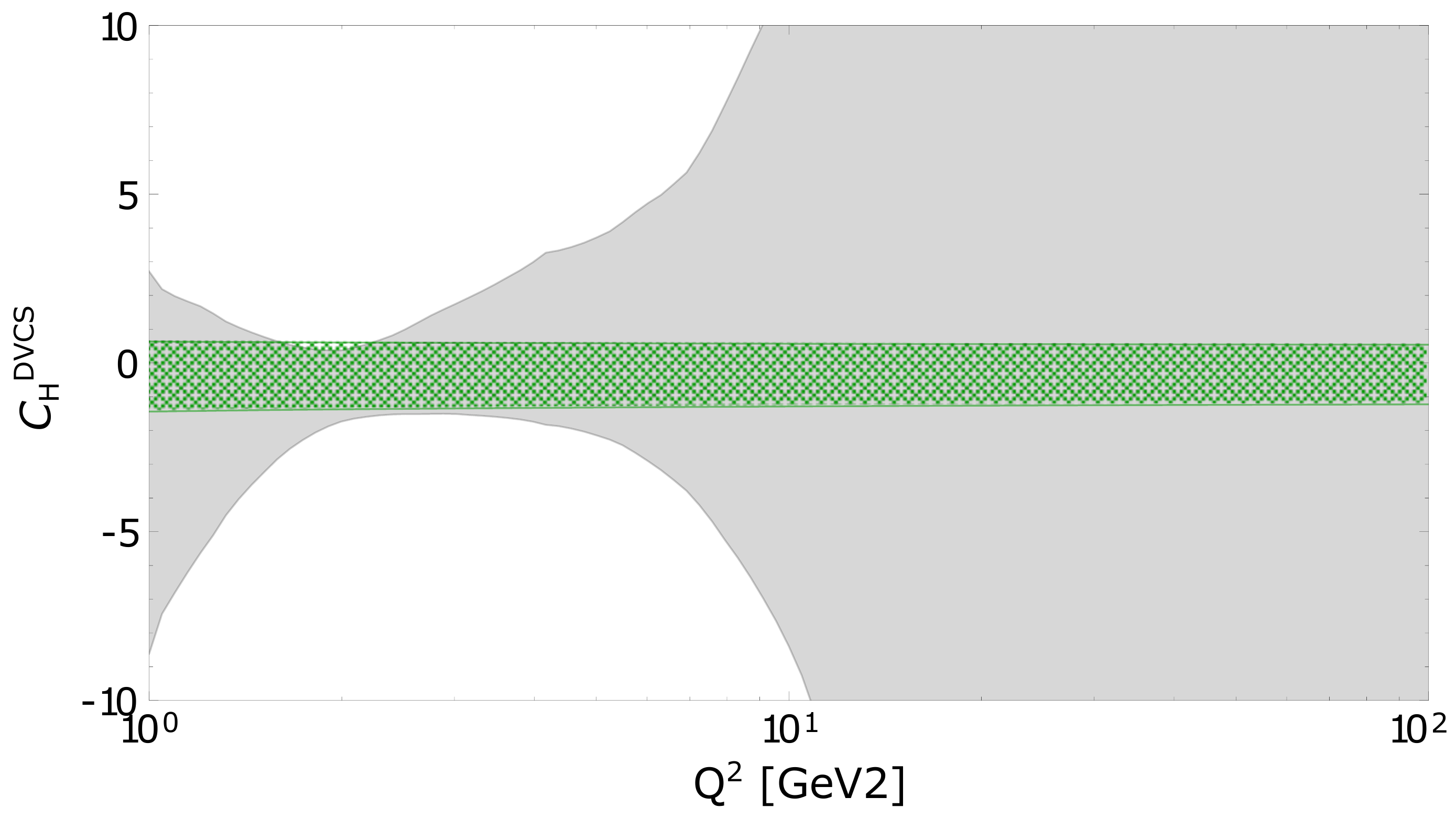}
\caption{Subtraction constant for the CFF $\mathcal{H}$ as a function of $\xi$ at $t = -0.3~\mathrm{GeV}^{2}$ and $Q^{2} = 2~\mathrm{GeV}^{2}$ (top), as a function of $-t$ at $\xi = 0.2$ and $Q^{2} = 2~\mathrm{GeV}^{2}$ (middle) and as a function of $Q^{2}$ at $\xi = 0.2$ and $t = -0.3~\mathrm{GeV}^{2}$ (bottom). The solid gray bands represent results of the neural network analysis of Ref.~\cite{Moutarde:2019tqa} while the green punctured bands represent the analysis based on a tripole Ansatz described in the text. All bands correspond to a $68 \%$ confidence level.}
\label{fig:results:sc}
\end{center}
\end{figure}

\begin{figure}[!ht]
\begin{center}
\includegraphics[width=\figWidth]{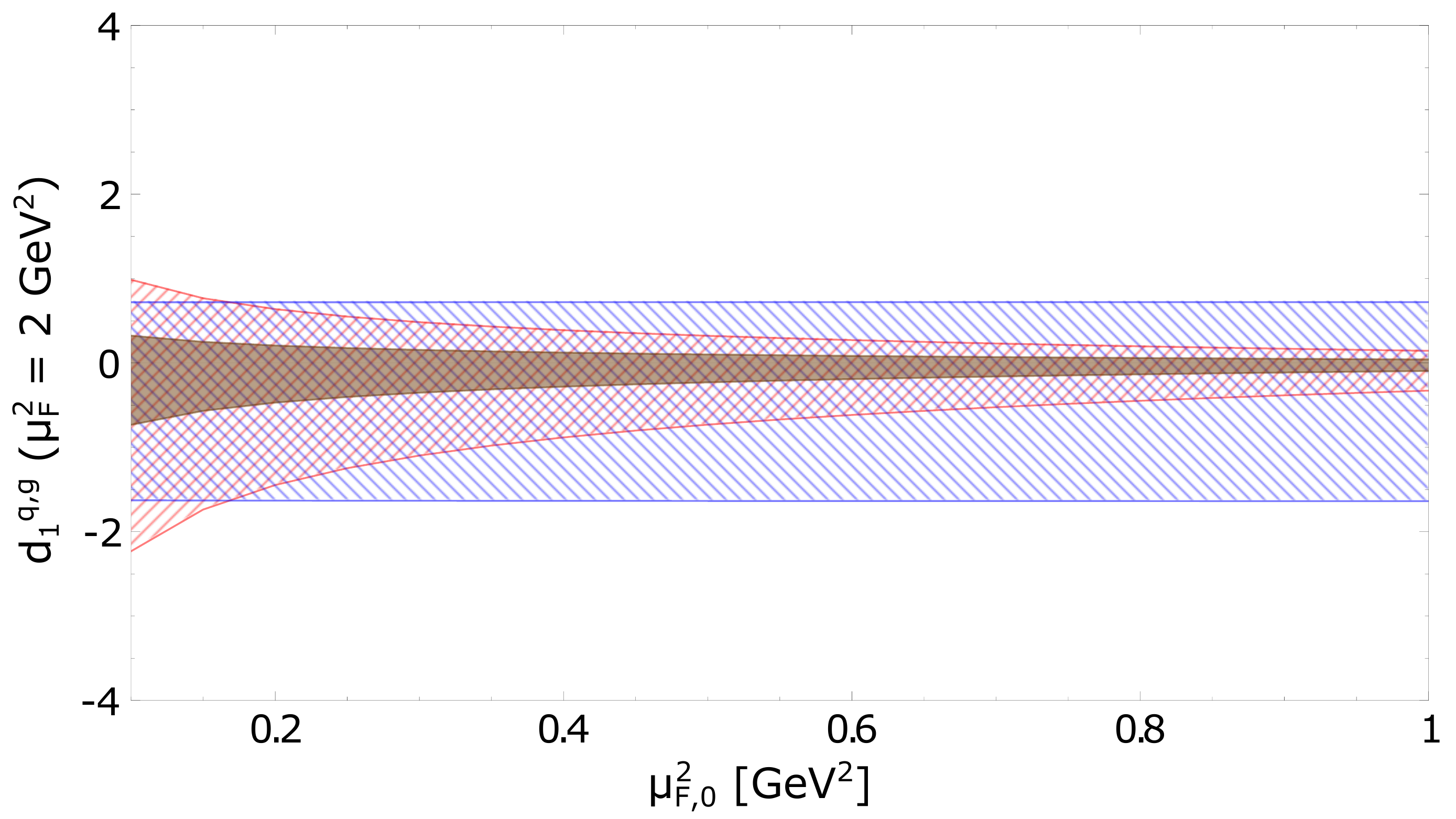}
\caption{Values of $d_{1}^{uds}(\MuF^2)$ for light quarks (blue backward-hatched band -- \textcolor{blue}{$\diagdown\diagdown$}), $d_{1}^{c}(\MuF^2)$ for charm quarks (brown solid band) and $d_{1}^{g}(\MuF^2)$ for gluons (red forward-hatched band -- \textcolor{red}{$\diagup\diagup$}) quoted at $\MuF^2 = 2~\mathrm{GeV}^2$ as a function of the initial factorisation scale squared $\mu_{\mathrm{F}, 0}^{2}$. The results for charm quarks are multiplied by $100$. All bands correspond to $68 \%$ confidence level.}
\label{fig:results:mu20}
\end{center}
\end{figure}

Means and uncertainties of the extracted parameters are estimated using Eqs. \eqref{eq:mean} and \eqref{eq:sigma}, respectively. The dependence of the subtraction constant on $\xi$, $t$ and $Q^{2}$ is shown in Fig. \ref{fig:results:sc} for both the tripole Ansatz and neural network replicas. One may notice that the uncertainties obtained with the tripole Ansatz are typically smaller than those obtained in the neural network analysis, which is expected and reflects the internal consistency of our approach. This is particularly visible in domains where the subtraction constant is strongly driven by the assumed functional form of the Ansatz, like in particular for large $|t|$. In Fig.~\ref{fig:results:sc} the difference between the confidence levels indicates the model uncertainty of the tripole Ansatz. The dependence of $d_{1}^{uds}(\MuF^2)$, $d_{1}^{g}(\MuF^2)$ and $d_{1}^{c}(\MuF^2)$ at $\MuF^2 = 2~\mathrm{GeV}^2$ on the choice of the initial factorisation scale, $\mu_{\mathrm{F}, 0}^{2}$ is shown in Fig.~\ref{fig:results:mu20}. As one may notice, the impact of this choice on the extracted value of $d_{1}^{uds}(Q^{2} = 2~\mathrm{GeV}^{2})$ is negligible compared to the estimated statistical uncertainties. 

\begin{figure}[!ht]
\begin{center}
\includegraphics[width=\figWidth]{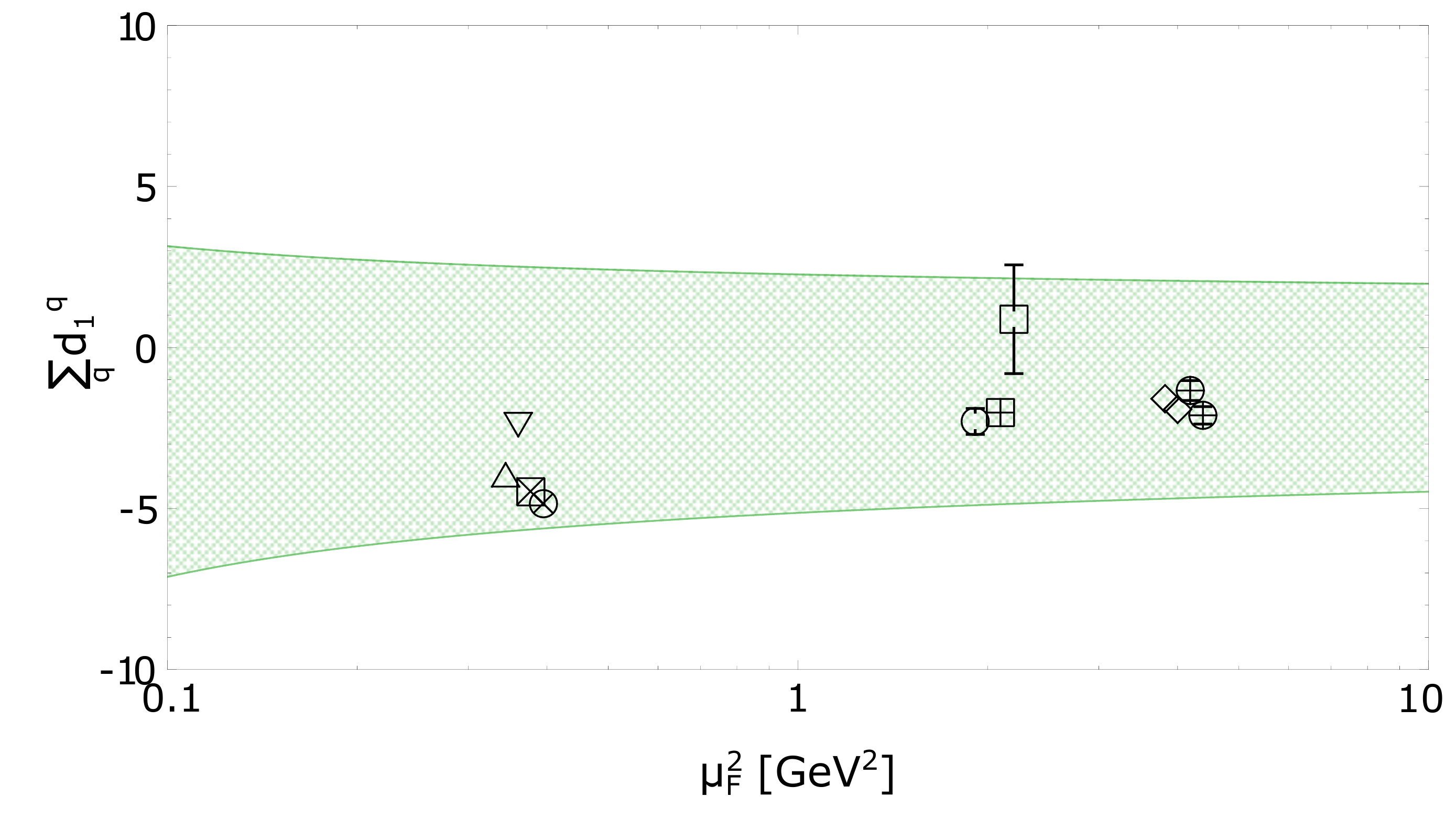}
\caption{The sum $\sum_{q} d_{1}^{q}(\MuF^{2})$ as a function of $\MuF^{2}$ for this study (green band) and other phenomenological and theoretical analyses. See Table \ref{tab:results:compilation} for the description of each data point, including the marker legend. For the sake of legibility some markers are artificially shifted by a small distance in the horizontal direction.}
\label{fig:results:compilation}
\end{center}
\end{figure}

Although error bars are too large to make a firm claim, we observe that a negative mean value of $d_{1}^{uds}(\MuF^2 = 2~\mathrm{GeV}^2)$ seems to be favoured by experimental data, as expected and observed in all stable systems \cite{Polyakov:2018zvc}. The results of other phenomenological and theoretical analyses, including lattice-QCD predictions, are collected in Tab.~\ref{tab:results:compilation}. Their comparison to our results is summarised Fig.~\ref{fig:results:compilation}.  

\begin{table*}[!ht]
\caption{Compilation of results for $\sum_{q} d_{1}^{q}(\MuF^{2})$ over a given number of quark flavours. Results 1) and 2) are originally only given for the DVCS subtraction constant considering three light flavours, and here are scaled by a factor $9/8$ following Eq. \eqref{eq:subtraction-vs-d1-uds}. The dispersive evaluations 3) differ by the input pion PDF. The lattice results 9) are originally only given for the EMT form factor $C_{u}(0, \MuF^{2})+C_{d}(0, \MuF^{2})$, and here are scaled by a factor $5$ following Eq. \eqref{eq:d1q-to-quark-GFF}. Both differ by the extrapolation of lattice data to the chiral limit. The scale associated to all results coming from $\chi$QSM is assumed to be $0.6~\mathrm{GeV}$, which is the natural scale for this type of models as argued in Ref.~\cite{Schweitzer:2002nm}. The same scale is associated to the Skyrme model.} 
\begin{center}
\begin{tabular}{@{}ccccccc@{}}
\toprule
\multirow{2}{*}{No.} & Marker         & \multirow{2}{*}{$\displaystyle\sum_{q} d_{1}^{q}(\MuF^{2})$} & $\MuF^{2}$ & \# of   & \multirow{2}{*}{Type} & \multirow{2}{*}{Ref.} \\
                     & in Fig. \ref{fig:results:compilation}  &                              & in $\mathrm{GeV}^2$ & flavours &      &      \\ 
\midrule
1   & $\ocircle$         & $-2.30 \pm 0.16 \pm 0.37$ & 2.0  & $3$ & from experimental data      & \cite{Burkert:2018bqq}          \\
2   & $\square$          & $0.88 \pm 1.69$           & 2.2  & $2$ & from experimental data      & \cite{Kumericki:2019ddg}          \\
3   & $\Diamond$         & $-1.59$                   & $4$    & $2$ & $t$-channel saturated model & \cite{Pasquini:2014vua}          \\
    &                    & $-1.92$                   & $4$    & $2$ & $t$-channel saturated model & \cite{Pasquini:2014vua}          \\
4   & $\triangle$        & $-4$                      & $0.36$ & $3$ & $\chi$QSM            & \cite{Goeke:2001tz}          \\
5   & $\bigtriangledown$ & $-2.35$                   & $0.36$ & $2$ & $\chi$QSM           & \cite{Goeke:2007fp}          \\
6   & $\boxtimes$        & $-4.48$                   & $0.36$ & $2$ & Skyrme model                & \cite{Cebulla:2007ei}          \\
7   & $\boxplus$         & $-2.02$                   & $2$    & $3$ & LFWF model                  & \cite{Muller:2014tqa}          \\
8   & $\otimes$          & $-4.85$                   & $0.36$ & $2$ & $\chi$QSM            & \cite{Wakamatsu:2007uc}          \\
9   & $\oplus$           & $-1.34 \pm 0.31$          & $4$    & $2$ & lattice QCD ($\overline{\mathrm{MS}}$)      & \cite{Hagler:2007xi}          \\
    &                    & $-2.11 \pm 0.27$          & $4$    & $2$ & lattice QCD ($\overline{\mathrm{MS}}$)      & \cite{Hagler:2007xi}          \\ \bottomrule
\end{tabular}
\end{center}
\label{tab:results:compilation}
\end{table*}

\subsection{Relaxing constraints on the fitting Ansatz}
\label{sec:relaxing-constraints}

We also tried to fit more than one parameter in the multipole Ansatz, namely $\{d_{3}^{uds}(\MuF^2), d_{1}^{g}(\MuF^2), \Lambda, \alpha\}$. The attempts at extracting two parameters at the same time are summarised in Tab.~\ref{tab:result:corr}. We first observe that all five extracted values of $d_{1}^{uds}(\MuF^2)$ (from Tabs.~\ref{tab:result:def} and \ref{tab:result:corr}) are consistent and compatible with zero. The ratio of the extracted mean to standard deviation vary over an order of magnitude, between $0.04$ and $0.5$ in all five fitting scenarios, confirming the difficulty of extracting a statistically significant value of $d_{1}^{uds}(\MuF^2)$ from existing DVCS data.

\begin{table}[!ht]
\caption{Results for four scenarios when $d_{1}^{uds}(\MuF^2)$ and one other parameter in $\{d_{3}^{uds}(\MuF^2), d_{1}^{g}(\MuF^2), \Lambda, \alpha\}$ are extracted from experimental data. This pair of parameters is noted $Y$. The values of are quoted at $\MuF^2 = 2~\mathrm{GeV}^2$. The last column indicates ranges in which parameters are allowed to vary in the fit procedure. If the range is not specified, the corresponding parameter is allowed to vary between $-\infty$ and $\infty$.}
\label{tab:result:corr}
\begin{center}
\begin{tabular}{@{}ccccc@{}}
\toprule
\multirow{2}{*}{No.} & \multicolumn{1}{c}{Fitted} & \multirow{2}{*}{Value} & \multicolumn{2}{c}{Allowed range} \\
& parameters $Y$ & & \multicolumn{1}{c}{min} & \multicolumn{1}{c}{max} \\\midrule
1 & $d_1^{uds}(\MuF^2)$ & $-0.7 \pm 1.2$ && \\
  & $d_1^{g}(\MuF^2)$ & $51 \pm 111$ && \\ \\
2 & $d_1^{uds}(\MuF^2)$ & $11 \pm 25$ && \\
  & $d_3^{uds}(\MuF^2)$ & $-11 \pm 26$ && \\ \\
3 & $d_1^{uds}(\MuF^2)$ & $-0.4 \pm 2.4$ && \\
  & $\Lambda/\mathrm{GeV}$ & $1.17 \pm 0.80$ &$0$&$2$ \\ \\
4 & $d_1^{uds}(\MuF^2)$ & $0.0 \pm 3.8$ && \\
  & $\alpha$ & $3.3 \pm 3.9$ &$0$&$10$ \\ \bottomrule
\end{tabular}
\end{center}
\end{table} 

The extraction introducing $d_{3}^{uds}(\MuF^2)$ as a free parameter has a somewhat different status from those proposing $\Lambda$ or $\alpha$ as their free parameters. While fixing $\Lambda$ and $\alpha$ imposes a parametric form to capture the $t$-dependence of experimental data, fitting only $d_1^{uds}(\MuF^2)$ represents a strong assumption on the behavior of the $(d_n^q)_{n \, \textrm{odd}}$ series. As mentioned in Sec.~\ref{sec:model}, this series has no expected reason to converge quickly and even less to stop after its first term. If a reasonable model can guide the choice of a specific $t$-dependence, \textit{a contrario} neglecting $d_{3}^{uds}(\MuF^2), d_{5}^{uds}(\MuF^2), \ldots$ in the extraction may leave a major uncontrolled source of systematic uncertainties.

In this respect the four scenarios restricting the series (\ref{eq:Dq}) to its first term provide us values of $d_{1}^{uds}(\MuF^2)$ in good agreement. The multipole parameters $\Lambda$ and $\alpha$ are still subject to large uncertainties. However, we observe a dramatic change in the estimated mean values and standard deviations between the fits extracting $d_{1}^{uds}(\MuF^2)$ only on the one hand, and $d_{1}^{uds}(\MuF^2)$ and $d_{3}^{uds}(\MuF^2)$ simultaneously on the other hand. This signals a possible influence of higher-order terms in the series (\ref{eq:Dq}) and raises the question of determining the terms of the series (\ref{eq:Dq}) from the values of the subtraction constant and the specific $Q^2$-dependence of each term.

\begin{figure}[!ht]
\begin{center}
\includegraphics[width=0.7\figWidth]{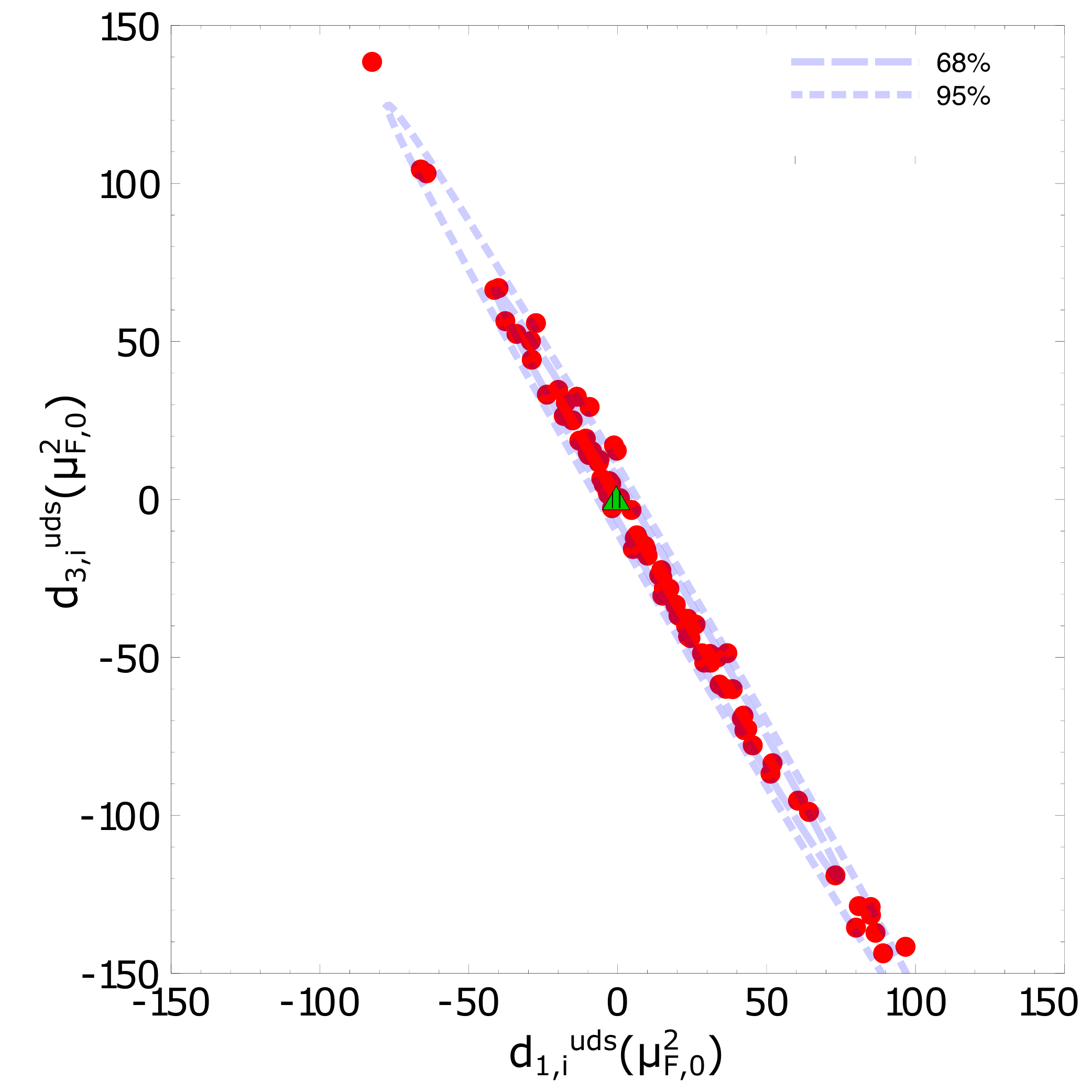}
\caption{Correlation between $d_1^{uds}(\MuFRef^2)$ and $d_3^{uds}(\MuFRef^2)$ obtained in the fitting scenario where both parameters are released at the same time. Each point corresponds to a single replica used in the extraction. Confidence levels of $68\%$ and $95\%$ are denoted by dashed and dotted curves, respectively. Pearson's correlation coefficient is estimated to be $-0.997$. The result of the extraction where only $d_1^{uds}(\MuFRef^2)$ is fitted (and where $d_3^{uds}(\MuFRef^2) = 0$) is denoted by the green triangle.}
\label{fig:results:corr_d1_d3}
\end{center}
\end{figure}

Separating $d_1$ from the higher-order terms $d_3$, $d_5$, \ldots consists in carefully identifying in the data their own logarithmic $Q^2$-dependence, driven by distinct anomalous dimensions. This requires a large lever arm in $Q^2$ and very accurate data, making this study best adapted to collider settings like EIC or EIcC. We observe that the joint fit of $d_{1}^{uds}(\MuF^2)$ and $d_{3}^{uds}(\MuF^2)$ to existing measurements exhibits a large correlation between these two parameters with a Pearson's correlation coefficient of $-0.997$, as shown in Fig.~\ref{fig:results:corr_d1_d3}. This plot summarises the difference between fitting $d_{1}^{uds}$ only, or $d_{1}^{uds}$, $d_{3}^{uds}$ simultaneously and depicts the aforementioned change between these two fitting scenarios. It also shows that the accuracy of existing data cannot exclude the case where higher-order terms in the series (\ref{eq:Dq}) decrease slowly.

\subsection{Discussion}
\label{sec:discussion}

Even if our results tend to favour a negative mean value for the first coefficient of the Gegenbauer expansion of the $D$-term $\sum_q d_{1}^{q}$ at $t=0$, we stress again that the estimated standard deviation makes it statistically compatible with 0. 

While we used a non-parametric global fit of almost all existing DVCS experimental data to obtain this result, the authors of Ref.~\cite{Burkert:2018bqq} reported a value of the same coefficient statistically incompatible with 0 (see Table~\ref{tab:results:compilation}) when using only a subset of the existing DVCS measurements (beam-spin asymmetries and cross sections published by CLAS in 2008 \cite{Girod:2007aa} and 2015 \cite{Jo:2015ema}) and a parametric fit of CFFs inspired by the KM model, see Ref.~\cite{Kumericki:2016ehc} and references therein. Extracting an apparently \emph{more accurate estimate} of $d_1$ from \emph{less experimental data} may come as a surprise, but it merely reflects the impact of systematic assumptions made in this type of studies and discussed in previous sections. 

The conclusions of our nominal fitting scenario are in both quantitative and qualitative agreement with those of Ref.~\cite{Kumericki:2019ddg}, which also rely on a neural network extraction of CFFs: $\sum_q d_{1}^{q}$ is found compatible with 0, with a mean to standard deviation ratio of about 0.5 (see Table \ref{tab:results:compilation}).  
The difference between the results reported here and in Ref.~\cite{Kumericki:2019ddg} that is observed in Fig.~\ref{fig:results:compilation} may be explained by a different extrapolation to $t=0$. We conducted here a careful study to evaluate the systematic uncertainties introduced by some classical assumptions, and we found no options allowing for a more accurate determination of $d_{1}^{uds}$ from the DVCS experimental data that have been collected so far. 

Our result being consistent with 0 within error bars, we will not perform the extraction of a pressure anisotropy profile from experimental data along the lines of Sect.~\ref{sec:emt}. In the same spirit, we do not attempt to evaluate the proton mechanical radius, which, from Eq.~\eqref{eq:def-mechanical-radius}, is essentially the second moment of this pressure profile, and would come compatible with 0 as well.

We nevertheless make a simple remark: if we had performed such an extraction, the distribution of pressure forces would be encoded in the $t$-dependence of $d_1$, which in current studies is described by the classical multipole Ansatz \eqref{eq:tripol}. Stated differently, the \emph{qualitative shape} of the pressure profile is selected before any fitting is realised, since it is merely the Fourier transform of a multipole. In this context the shape of the pressure profile is a model assumption, and this shape may vary with the choice of the multipole parameters $\Lambda$ and $\alpha$, and the overall normalisation $d_1(t=0)$.

\begin{figure}[!ht]
\begin{center}
\includegraphics[width=\figWidth]{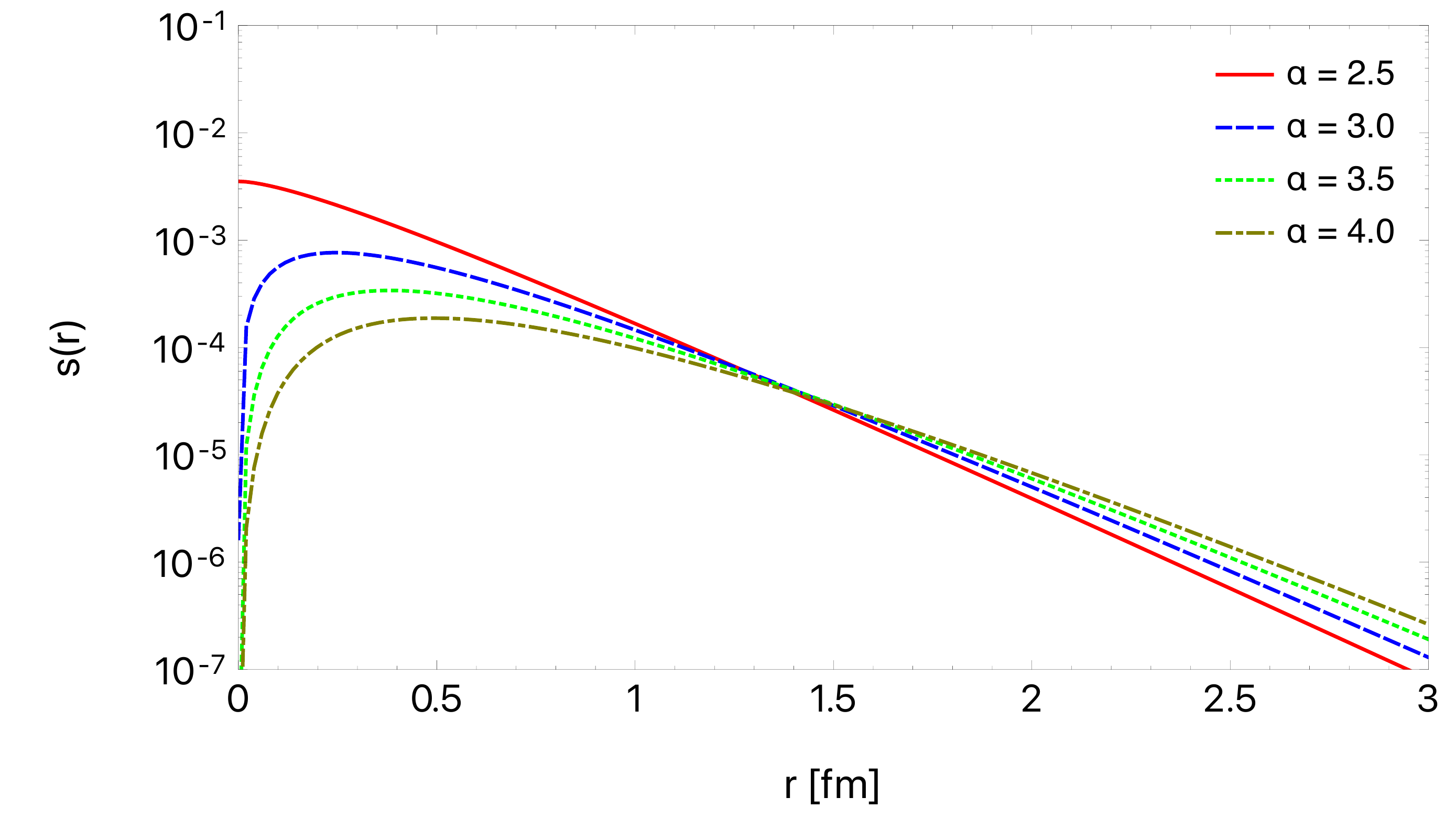}
\includegraphics[width=\figWidth]{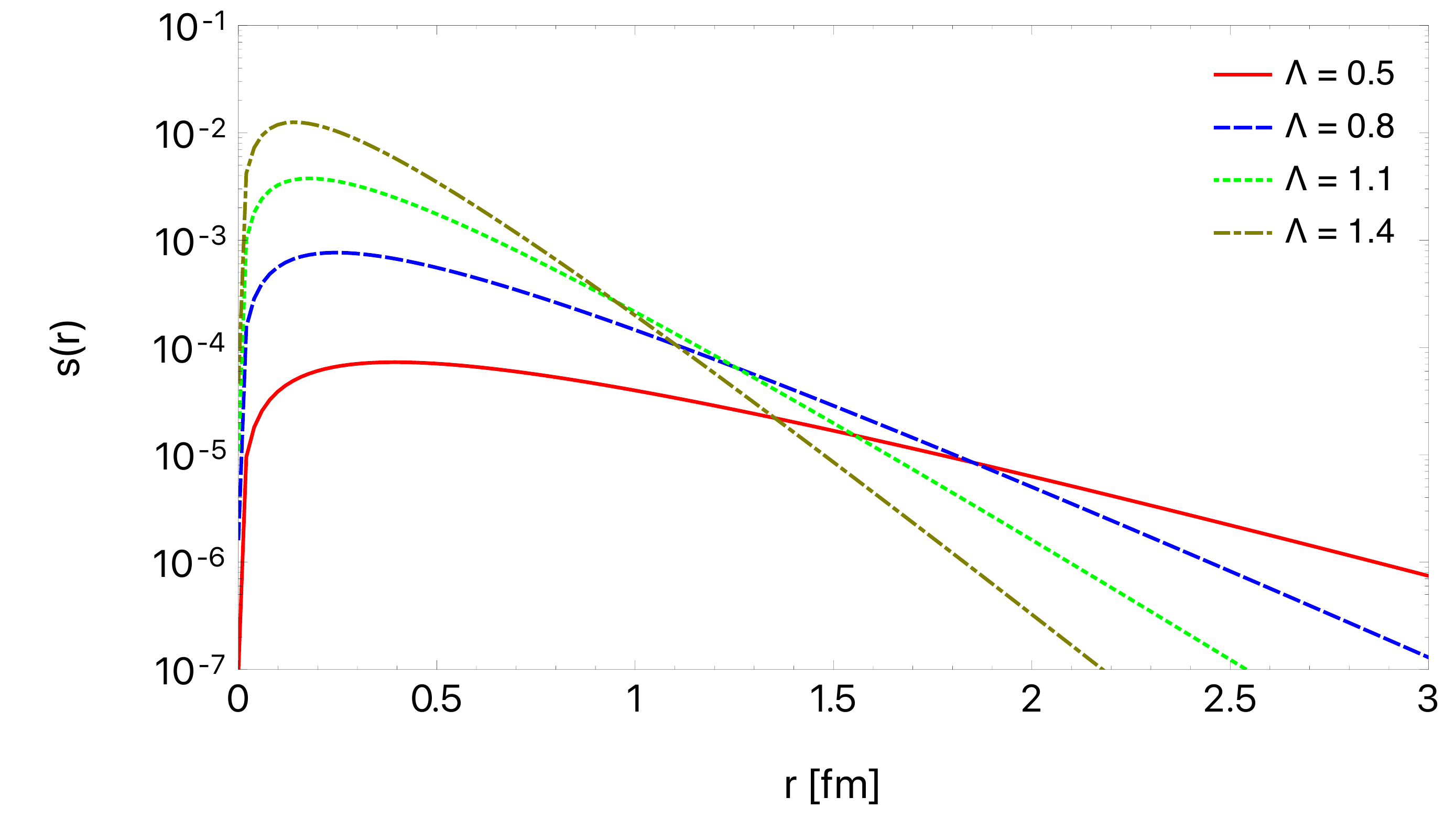}
\caption{Profiles of pressure anisotropies for a single parton, see Eqs. \eqref{eq:pressure_profile_q} and \eqref{eq:pressure_profile_g}, using the multipole Ansatz of $d^{q,g}_{1}(t)$, see Eq.~\eqref{eq:tripol}, for various values of the parameters $\alpha$ and $\Lambda$. The upper plot corresponds to $\Lambda = 0.8~\mathrm{GeV}$ and various values of $\alpha$, while the lower one assumes $\alpha = 3$ and shows various values of $\Lambda$. In both cases the same normalisation factor is used: $d_{1}(0) = -3$.}
\label{fig:results:profiles}
\end{center}
\end{figure}

We use Eq.~\eqref{eq:pressure_profile_q} to point out the dependence of this profile on the assumed values of the tripole Ansatz parameters considering the results of Tab.~\ref{tab:result:corr}. Since we did not estimate $\Lambda$ and $\alpha$ simultaneously from the data, we do not know the correlation matrix of these parameters. Thus we performed an impact study by varying only one parameter and keeping the other fixed. This may allow us to explore a larger parametric space than otherwise permitted by DVCS data, but the exercise reported in Fig.~\ref{fig:results:profiles} demonstrates variations in the pressure profile over orders of magnitude. As a consequence, all phenomenological information about the distribution of pressure forces extracted to this day from current experimental data should be taken with great care. As emphasised above, future data from EIC or EIcC should improve these phenomenological results and clarify the situation.

%
%
\section{Summary}
\label{sec:summary}

This article summarises our analysis of the $d_{1}(\MuF^{2})$ coefficient extracted from the world experimental data for the DVCS process. This coefficient is of utmost importance for the understanding of the QCD energy-momentum tensor. It carries information on mechanical properties of partonic media, like shear stress and pressure. 

We carefully detailed the extraction procedure, and plainly exhibited a set of assumptions that we made and that have been used in other analyses of similar type until now. The underlying motivation is the assessment of what can quantitatively be said to this day about the proton mechanical properties using DVCS measurements. We hope it will be a useful starting point for future studies when new data become available. We tried to make this document as self-contained as possible, including all equations with consistent notations, covering the whole reasoning leading to the extraction of pressure distributions, from proton structure to DVCS measurements. In particular, the evolution equations of the $d_{n}^{q,g}(\MuF^{2})$ coefficients are presented in a complete form and are used to estimate the contributions coming from gluons and charm quarks. 

The comparison between analyses where either artificial neural networks or parametric functional forms are used to describe CFFs clearly demonstrates the model dependence of the latter approach, which results in considerably smaller uncertainty for the extracted $d_{1}^{uds}(\MuF^{2})$ coefficient, making it no longer compatible with 0. In such a case the systematic uncertainty associated to the choice of a specific functional form is difficult to evaluate and is left as an unknown. The extraction of $d_{1}^{uds}(\MuF^{2})$ independently from elements proportional to higher terms in the $D$-term Gegenbauer expansion, like $d_{3}^{uds}(\MuF^{2})$, cannot be performed at this moment. Removing the assumption regarding the symmetry of the quark coefficients $d_{n}^{uds}(\MuF^{2})$ most probably requires a multi-channel analysis.

Even if the link between the distribution of pressure forces in the proton and the DVCS subtraction constant is well-defined, and even if this link is subject to approximations that can be checked systematically, DVCS data do not allow yet a statistically significant extraction of these pressure forces. Moreover the present need of a prior knowledge of the $t$-dependence of the $D$-term drives the pressure profile. All phenomenological information about the distribution of pressure forces extracted to this day from current experimental data should be taken with great care. Our analysis establishes the need for more precise data and for an extension of the covered kinematic domain. This can be achieved by future experiments to be conducted at JLab, CERN, EIC and EIcC facilities.

\begin{acknowledgements}
The authors thank F.-X.~Girod and K.~Kumeri\v{c}ki for valuable discussions. This project was supported by the European Union's Horizon 2020 research and innovation programme under grant agreement No 824093, the Grant No. 2017/26/M/ST2/01074 of the National Science Centre, Poland and was realised with the support of the French Government scholarships programme. The project is co-financed by the Polish National Agency for Academic Exchange and by the COPIN-IN2P3 Agreement. The computing resources of {\'S}wierk Computing Centre, Poland are greatly acknowledged. 
We acknowledge further financial support from the Agence Nationale de la Recherche (ANR) under the projects No. ANR-18-ERC1-002 and ANR-16-CE31-0019, the P2IO LabEx (ANR-10-LABX-0038) in the framework ``Investissements d'Avenir'' (ANR-11-IDEX-0003-01) managed by the ANR, and the CEA-Enhanced Eurotalents Program co-founded by FP7 Marie Sk\l odowska-Curie COFUND Program (Grant agreement No. 600382).
\end{acknowledgements}

\appendix
\section{Open source code}
\label{appendix:opensource_code}

Replicas used in this analysis representing the subtraction constant parameterised with the tripole Ansatz  are available in PARTONS framework \cite{Berthou:2015oaw} as the module \texttt{GPDSubtractionConstantDLMSTW21}.
The code of this framework is open source and can be found online at \url{https://drf-gitlab.cea.fr/partons/core/partons} on version 3 of the GPL (GPLv3).

\bibliographystyle{spphys}
\bibliography{bibliography}

\end{document}